%% file: template.tex
\documentclass[journal]{vgtc}                

\ifpdf
  \pdfoutput=1\relax                   
  \usepackage{graphicx}                
  \DeclareGraphicsExtensions{.pdf,.png,.jpg,.jpeg} 
\else
  \usepackage{graphicx}                
  \DeclareGraphicsExtensions{.eps}     
\fi%

\graphicspath{{figures/}{pictures/}{images/}{./}} 

\usepackage{epstopdf}
\usepackage{microtype}
\PassOptionsToPackage{warn}{textcomp}
\usepackage{textcomp}
\usepackage{mathptmx}
\usepackage{times}

\usepackage{cite}
\usepackage{tabu}
\usepackage{booktabs}
\usepackage{wrapfig}

\usepackage{amssymb}
\usepackage{paralist}
\usepackage{xfrac}
\usepackage{url}
\usepackage{algorithm}
\usepackage{algpseudocode}
\usepackage{amsmath}
\usepackage{multirow}
\usepackage{xcolor}
\usepackage{todonotes}
\usepackage[shortlabels]{enumitem}
\usepackage[xcolor=dvipdf, final]{changes}

\definecolor{mygreen}{rgb}{0.17, 0.55, 0.05}
\definecolor{myred}{rgb}{0.85, 0.17, 0.05}
\definechangesauthor[name=JiachengPan, color=mygreen]{pan}
\definechangesauthor[name=XiaodongZhao, color=red]{zhao}
\definechangesauthor[name=Jian Chen, color=blue]{jc}
\setdeletedmarkup{\color{gray}{\sout{#1}}}
\setaddedmarkup{\color{mygreen}{#1}}
\setauthormarkup{}

\onlineid{1069}

\vgtccategory{Research}
\vgtcpapertype{Algorithm/Technique}

\title{Exemplar-based Layout Fine-tuning for Node-link Diagrams}
\author{Jiacheng Pan, Wei Chen, Xiaodong Zhao, Shuyue Zhou, Wei Zeng, Minfeng Zhu, \\ Jian Chen, Siwei Fu, and Yingcai Wu}
\authorfooter{
    \item Jiacheng Pan, Xiaodong Zhao, Shuyue Zhou, Minfeng Zhu, Yingcai Wu, and Wei Chen are with the State Key Lab of CAD\&CG, Zhejiang University, China. E-mail: \{panjiacheng, zhaoxiaodong, zhoushuyue, minfeng\_zhu, ycwu\}@zju.edu.cn, chenwei@cad.zju.edu.cn.
    \\
    Yingcai Wu is also with Zhejiang Lab, China.\\
    Wei Chen and Yingcai Wu are the corresponding authors.
    
    \item Wei Zeng is with Shenzhen Institutes of Advanced Technology, Chinese Academy of Sciences, China. E-mail: wei.zeng@siat.ac.cn.
    
    \item Jian Chen is with Ohio State University, USA. E-mail: chen.8028@osu.edu.
    
    \item Siwei Fu is with Zhejiang Lab, China. E-mail: fusiwei339@gmail.com.
}

\shortauthortitle{Pan \MakeLowercase{\textit{et al.}}: Exemplar-based Layout Fine-tuning for Node-link Diagrams}

\abstract{We design and evaluate a novel layout fine-tuning technique for node-link diagrams that facilitates exemplar-based adjustment of a group of substructures in batching mode. The key idea is to transfer user modifications on a local substructure to other substructures in the entire graph that are topologically similar to the exemplar. We first precompute a canonical representation for each substructure with node embedding techniques and then use it for on-the-fly substructure retrieval. We design and develop a light-weight interactive system to enable intuitive adjustment, modification transfer, and visual graph exploration. \replaced[id=pan]{\textcolor{black}{We also report some results of quantitative comparisons, three case studies, and a within-participant user study.}}{Experimental results and case studies demonstrate that our approach improves readability and performance over existing layout editing schemes.}}


\CCScatlist{ 
 \CCScat{K.6.1}{Management of Computing and Information Systems}%
{Project and People Management}{Life Cycle};
 \CCScat{K.7.m}{The Computing Profession}{Miscellaneous}{Ethics}
}

\teaser{
  \centering
    \includegraphics[width=16cm]{picture/finan512}
    \caption{Case study with the Finan512 dataset~\cite{10.1145/2049662.2049663} of 74,752 nodes and 261,120 edges: (a) in the node-link diagram generated with FM$^3$~\cite{hachul2004drawing}, we specify an exemplar (in red) and several similar substructures (in blue) are retrieved with $k = 200$, $min=10$, $max=100$, and $\epsilon=0.95$; (b) the exemplar; (c) five most similar (1-5) and five least similar (6-10) retrieved substructures; (d) we specify 14 substructures around the exemplar; (e) the modified exemplar; (f) modifications are transferred to 10 retrieved substructures; (g) modifications are transferred to substructures around the exemplar. All modified substructures are merged into the entire graph. \added{(h-k) Readability before (orange) and after (purple) modification transfer measured by four readability criteria (from top to bottom: crosslessness, minimum-angle metric, edge-length variant, and shape-based metric); error bars depict 95\% confidence intervals.}}
    \label{fig:Finan512}
}

\vgtcinsertpkg

\begin{document}

\input{chapter/1.Introduction.tex}
\input{chapter/2.RelatedWork.tex}
\input{chapter/3.Approach.tex}
\input{chapter/4.Algorithm.tex}
\input{chapter/5.Experiments.tex}
\input{chapter/6.Discussion.tex}
\input{chapter/7.Conclusion.tex}

\acknowledgments{
We wish to thank all the anonymous reviewers for their thorough and constructive comments. We also thank the participants for their time and efforts. 
This work is partially supported by National Natural Science Foundation of China (61772456, 61761136020), NSFC (61761136020), NSFC-Zhejiang Joint Fund for the Integration of Industrialization and Informatization (U1609217), and Zhejiang Provincial Natural Science Foundation (LR18F020001). J. Chen is partially supported by National Science Foundation NSF OAC-1945347, NSF DBI-1260795, NSF IIS-1302755, CNS-1531491, and NIST MSE-70NANB13H181. Any opinions, findings, and conclusions or recommendations expressed in this material are those of the authors and do not necessarily reflect the views of  the National Science Foundation of China (NSFC), National Institute of Standards and Technology (NIST) or the National Science Foundation (NSF).
}

\bibliographystyle{abbrv}
\bibliography{template}
\end{document}

%% file: chapter/1.Introduction.tex
\firstsection{Introduction}
\maketitle

\deleted{Node-link diagrams and matrix-based views are two main visualization methods for graph data. For large-scale graphs, node-link diagrams are preferable because they are capable of explicitly showing the topology (or connectivity) among a large number of nodes{~\cite{ghoniem2004comparison}}.}
Generating appropriate layouts of graph data has been a major research topic over the past decades, as witnessed by the extensive literature~\cite{DBLP:journals/ivs/CheongS20, DBLP:reference/crc/2013gd, DBLP:journals/csur/DiazPS02, DBLP:journals/ivs/GibsonFV13, DBLP:journals/tvcg/HermanMM00}.
\added{Among many solutions, node-link diagrams are widely used, because they reveal topology and connectivities{~\cite{ghoniem2004comparison}}.}
When the number of nodes and edges increases, algorithms aimed at both computational speed and readability are valuable. New force-directed layout algorithms have harnessed data features to layout a large set of nodes and edges effectively~\cite{jacomy2014forceatlas2}.
However, additional layout optimizations or manual modifications are typically required to improve readability~\cite{DBLP:journals/tvcg/WangWSZLFSDC18}.

The aesthetics of a graph layout is often subjective and may vary with user preferences.
Modern rule-based graph layout methods~\cite{dwyer2017cola, DBLP:journals/cgf/HoffswellBH18, DBLP:journals/tvcg/KiefferDMW16, DBLP:journals/tvcg/WangWSZLFSDC18} have successfully integrated users' preferences into layout.
Nevertheless, such state-of-the-art solutions for interactive fine-tuning of node-link diagrams work only for either 
dragging individual nodes 
or the entire diagram (e.g., fisheye)~\cite{DBLP:journals/tvcg/CoheLBEL16, DBLP:conf/chi/DuCLXT17}. 
Interaction techniques have enabled
graph exploration but not layout modification~\cite{DBLP:journals/tvcg/ChenGHPNXZ19, DBLP:conf/vinci/YanCZ19}.
In particular, fine-tuning of node positions in a layout has to be manually performed, which is laborious and time-consuming.

We design an \textit{interactive exemplar-based tuning algorithm} for displaying node-link diagrams in which \textit{exemplar} is a local substructure of the underlying graph specified by users, following the technique proposed in~\cite{DBLP:journals/tvcg/ChenGHPNXZ19} (Figure~\ref{fig:Finan512}b).
The key to our solution is first to find topologically similar structures to the user-chosen exemplar and then to morph these structures automatically into a user-defined layout before embedding them in the original graph.
Transferring the user's input has two main challenges.
First, substructures in a large node-link diagram can have distinctive topologies. Identifying similar ones from the entire diagram and constructing the correspondences between the two substructures is a nontrivial task. 
Second, mapping the dynamic
change of one exemplar to another requires solving a two-dimensional substructure transformation.
Our solution to these challenges has three main components: \textit{representation}, 
\textit{retrieval}, and \textit{morphing of substructures}, designed to efficiently fine-tune substructures containing a group of user-specified nodes and edges. 
Compared with \replaced{the baseline method (manual node dragging)}{existing layout editing schemes~\cite{DBLP:conf/icwsm/BastianHJ09, shannon2003cytoscape}}, our approach facilitates fast specifications of substructures and local layout fine-tuning based on users' preferences.

This paper makes the following contributions:
\begin{compactitem}
    \item A novel layout fine-tuning method that can simultaneously adjust layouts of multiple similar substructures to user preferences;
    \item An efficient modification-transfer algorithm that can transfer fine-tuned results of an exemplar substructure to other topologically similar substructures;
    \item A set of quantitative and qualitative experiments that evaluate the efficiency of our approach.
\end{compactitem}

\deleted{The rest of this paper is organized as follows. Section 2 presents related works. A novel modification transfer algorithm is described in Section 3. Section 4 explains the entire layout fine-tuning approach and visual interface. Experiments and evaluation are given in Section 5. Section 6 concludes this paper.}

%% file: chapter/2.RelatedWork.tex
\section{Related Work}\label{sec:relatedwork}
\replaced{We review two related areas: graph visualization techniques and interaction techniques.}
{Node-link diagram is the most popular approach in graph drawing~\cite{DBLP:conf/dac/FiskI65}. Various solutions have been proposed for node placement.
Below we only elaborate a few of them. For a comprehensive survey, please refer to~\cite{DBLP:journals/ivs/CheongS20, DBLP:journals/csur/DiazPS02, DBLP:journals/ivs/GibsonFV13, DBLP:journals/tvcg/HermanMM00}.}

\subsection{Graph Visualization}
\replaced{
Two-dimensional (2D) graph drawing methods have been broadly reported in textbooks and surveys~\cite{DBLP:reference/crc/2013gd, DBLP:journals/ivs/CheongS20, DBLP:journals/ivs/GibsonFV13}.
Force-directed and related drawing methods are classified into three categories~\cite{DBLP:journals/ivs/GibsonFV13}:
force-directed methods,
dimension-reduction methods,
and multi-level methods.
Force-directed methods simulate physical forces on nodes and edges to layout graphs; many extensions exist, e.g.,
spring-embedded methods~\cite{eades1984heuristic, DBLP:journals/spe/FruchtermanR91, jacomy2014forceatlas2},
energy-based methods~\cite{DBLP:conf/gd/GansnerKN04, DBLP:journals/ipl/KamadaK89},
and probabilistic methods~\cite{journals/tog/DavidsonH96, DBLP:journals/swevo/KudelkaKRHS15}. Dimensionality-reduction methods aim to embed high-dimensional information (e.g., the shortest path length between two nodes) into a 2D space,
using methods such as multidimensional scaling~\cite{DBLP:conf/gd/BrandesP06}, 
self-organizing maps~\cite{DBLP:journals/ipl/BonabeauH98}, and t-SNE~\cite{DBLP:journals/cgf/KruigerRMKKT17}.
Multi-level methods focus on accelerating graph drawing using two main phases: coarsening (simplify a graph into several coarser graphs) and refinement (successively compute fine layouts from simple coarser graphs)~\cite{gansner2011multilevel, hachul2004drawing}.
Besides these generic layout algorithms for node-link diagrams, other methods aim to solve specific drawing problems.
For example, orthogonal layouts proposed in~\cite{DBLP:journals/tvcg/CoheLBEL16, DBLP:journals/tvcg/KiefferDMW16} improve readability of node-link diagrams of power-grids, software, and financial markets.
}{
Automatic graph drawing methods have been studied since 1967~\cite{DBLP:conf/dac/FiskI65}. These methods can be roughly classified into three categories: spring model, energy model and projection model.
The first force-directed algorithm~\cite{eades1984heuristic} based on spring model simulates physical forces on nodes and edges.
Date from then, the spring model is improved into the spring-electrical model in various ways~\cite{DBLP:conf/gd/DwyerMW06, eades2004navigating, frick1994fast, DBLP:conf/gd/GajerGK00, hu2005efficient}. The multi-level method is used widely to accelerate spring-electrical methods~\cite{gajer2000grip, gansner2011multilevel, hachul2004drawing}. 
The energy model-based methods formulate the layout problem as an optimization problem. 
The graph distance approximation~\cite{DBLP:journals/ipl/KamadaK89} and the incremental arrangement technique~\cite{DBLP:journals/tochi/Cohen97} can be leveraged to speed up the optimization process.
Then, stress majorization is employed in graph layout problem~\cite{DBLP:conf/gd/GansnerKN04}. 
Many constrained layout algorithms use stress majorization to formulate constraint problem~\cite{DBLP:journals/tvcg/KiefferDMW16, DBLP:journals/tvcg/WangWSZLFSDC18}.
The projection model-based methods usually embed nodes into a high-dimensional space. Then they use projection methods to project the high-dimensional vectors into low-dimensional space, i.e., 2-dimensional space. The first layout algorithm based on high-dimensional embedding (HDE) is proposed in 2004~\cite{DBLP:journals/jgaa/HarelK04}. 
Instead of using high-dimensional distance, some methods adopt graph distance in projection procedure~\cite{DBLP:journals/cgf/KruigerRMKKT17, DBLP:journals/corr/LuYC16}.}

\deleted{
\textbf{Constrained Layout}:
Constrained graph layout introduces rules for node placement, which is often employed to improve force-directed layouts. 
For example, PrEd is an iterative constrained drawing algorithm to reduce new edge crossings~\cite{DBLP:conf/gd/Bertault99}.
The algorithm can be further improved in both running time and drawing quality \cite{DBLP:journals/cgf/SimonettoAAB11}.
Constrained graph layout can also be applied to preserve the topology of force-directed layouts~\cite{DBLP:conf/gd/DwyerMW08}.
Energy minimization and stress majorization are also popular in constrained graph layout. Dig-CoLa~\cite{DBLP:conf/infovis/DwyerK05} automatically detects and places the parts of the graph that contain hierarchical information by stress majorization. 
Based on an authoring tool supporting constrained layout~\cite{DBLP:conf/gd/DwyerMW08a}, graph exploration and layout techniques have been proposed~\cite{DBLP:journals/tvcg/DwyerMSSWW08, DBLP:conf/gd/DwyerMW08}. {Cola.js}~\cite{dwyer2017cola} extends the authoring tool into a JavaScript library. SetCoLa~\cite{DBLP:journals/cgf/HoffswellBH18} further presents a domain-specific language with eight sets of predefined constraints. 
In terms of stress majorization, edge length is the most popular target to be optimized~\cite{DBLP:journals/tvcg/WangWSZLFSDC18, yuan2012intelligent}.
}

\replaced{
Unlike prior studies on layout algorithms,
our work focuses on interactive fine-tuning by capturing users' layout preferences 
through interaction. Our algorithms can potentially support personalized and fine-tuned layout of these current state-of-the-art graph visualizations.
}{
Although previous work has introduced rules to constrain automatic layouts, there are few approaches for interactive layout fine-tuning. Our work refines the graph layout by capturing the user's preference and intention through interactions.
}

\subsection{Interaction Techniques}
\deleted{Effective interactions are critical to graph analysis when the graph data is too complex or large to be visualized.}
We categorize interaction techniques into three levels: data-level, view-level, and 
encoding-level.

\textbf{Data-level interactions} focus on selecting the data for display.
The user can interact with the graphs to see similar structures.
A system developed in~\cite{von2009system} uses user-defined subgraph or motifs to reveal selected structures but these motifs were predefined and could not be modified by the users. 
Several systems~\cite{garbarino2016EGAS, wu2017graph, zhu2015pathrings} use PathRings to define motifs in biological pathways, but they do not find similar structures.
\added{
Novel machine learning solutions utilized in~\cite{DBLP:journals/tvcg/KwonCM18} measure the similarity between two graphs, but it is not feasible because it does not locate substructures.}
A structure-based recommendation approach~\cite{DBLP:journals/tvcg/ChenGHPNXZ19} detects similar substructures in a graph from user input and lets users subsequently interact with the detected structures.
\added{We adopt this approach to measure similarity, thus reducing user input; we subsequently introduce a new algorithm to further reduce users' repetitive and effortful node editing through a substructure transformation algorithm. 
}
\deleted{An alternative way of analyzing large-graph data is aggregation; the vertices of interest are recursively detected and merged according to the user-defined node attributes~\cite{wattenberg2006visual} or topological properties~\cite{archambault2007topolayout}
To support interactive graph visualization at different levels of detail, nodes and edges are aggregated in different levels using the kernel density estimation technique \cite{zinsmaier2012interactive,lampe2011interactive}.
Likewise, a hierarchical graph structure can give an overview of graph structure first and show more detail upon user interaction~\cite{elmqvist2009hierarchical}.}

\textbf{View-level interactions} mostly support graph navigation.
\deleted{Panning and zooming support multi-level explorations by navigating along edges of a selected node.}
Topology information can be exploited in browsing a large graph~\cite{moscovich2009topology}.
\deleted{Focus+context interactions are widely used to provide a detailed view of a subgraph while preserving an overview of the entire graph.}
Fisheyes enlarge the display space for items of user interest to improve readability.
For example,  
SchemeLens~\cite{DBLP:journals/tvcg/CoheLBEL16} reveals orthogonally laid-out diagrams.
And the structure-aware fisheye proposed in~\cite{DBLP:journals/tvcg/WangWZSFSCD19} reduces spatial and temporal distortions.
\added{Compared to these
solutions, our method supports the user's defined input to customize layout.}

\textbf{Encoding-level interactions} seek to manipulate the visual representation and layout of graph data. 
An appropriate layout can benefit analysis tasks~\cite{jusufi2013multivariate, major2018graphicle}. However, generating visually pleasing and
useful layouts for large graphs is still challenging.\deleted{, especially for large-scale graphs.
Graphs can be visualized by leveraging node-link diagrams~\cite{DBLP:journals/tvcg/ColVPDSN18} and adjacency matrices \cite{DBLP:journals/tvcg/HenryF06}.}
NodeTrix~\cite{henry2007nodetrix} combines two schemes to show inter-community relationships using a node-link diagram and intra-community relationships using the matrix representation. 
In many situations, analysts fine-tune the node positions.
An authoring tool proposed in~\cite{DBLP:conf/gd/DwyerMW08a} introduces continuous layout in response to user input.
A method that could integrate multiple graph layouts preserved topological structures in graphs by controlling the Euclidean distance between nodes of subgraphs \cite{yuan2012intelligent}.
Some constraint-based layout editing methods\replaced{~\cite{DBLP:conf/uist/RyallMS97, DBLP:journals/bmcbi/SchreiberDMW09, DBLP:journals/tvcg/WangWSZLFSDC18}}{~\cite{DBLP:journals/tvcg/WangWSZLFSDC18}} allow the user to edit and explore a layout with selected constraint rules. \added[id=pan]{However, these methods aim to draw a constraint-based layout, and could not edit a layout freely on nodes to reach a fine-tuned layout and incorporate users' preferences.
}

\deleted{However, it is still too time-consuming to fine-tune graph layout by dragging nodes of a large graph. Our approach addresses this problem using exemplar-based scheme.}

%% file: chapter/3.Approach.tex
\section{\replaced{Layout Fine-Tuning: Workflow and Interface}{Layout Fine-tuning For Node-Link Diagrams}}
\begin{figure}[!tp] 
    \centering
    \setlength{\belowcaptionskip}{-5pt}
    \includegraphics[width=1\columnwidth]{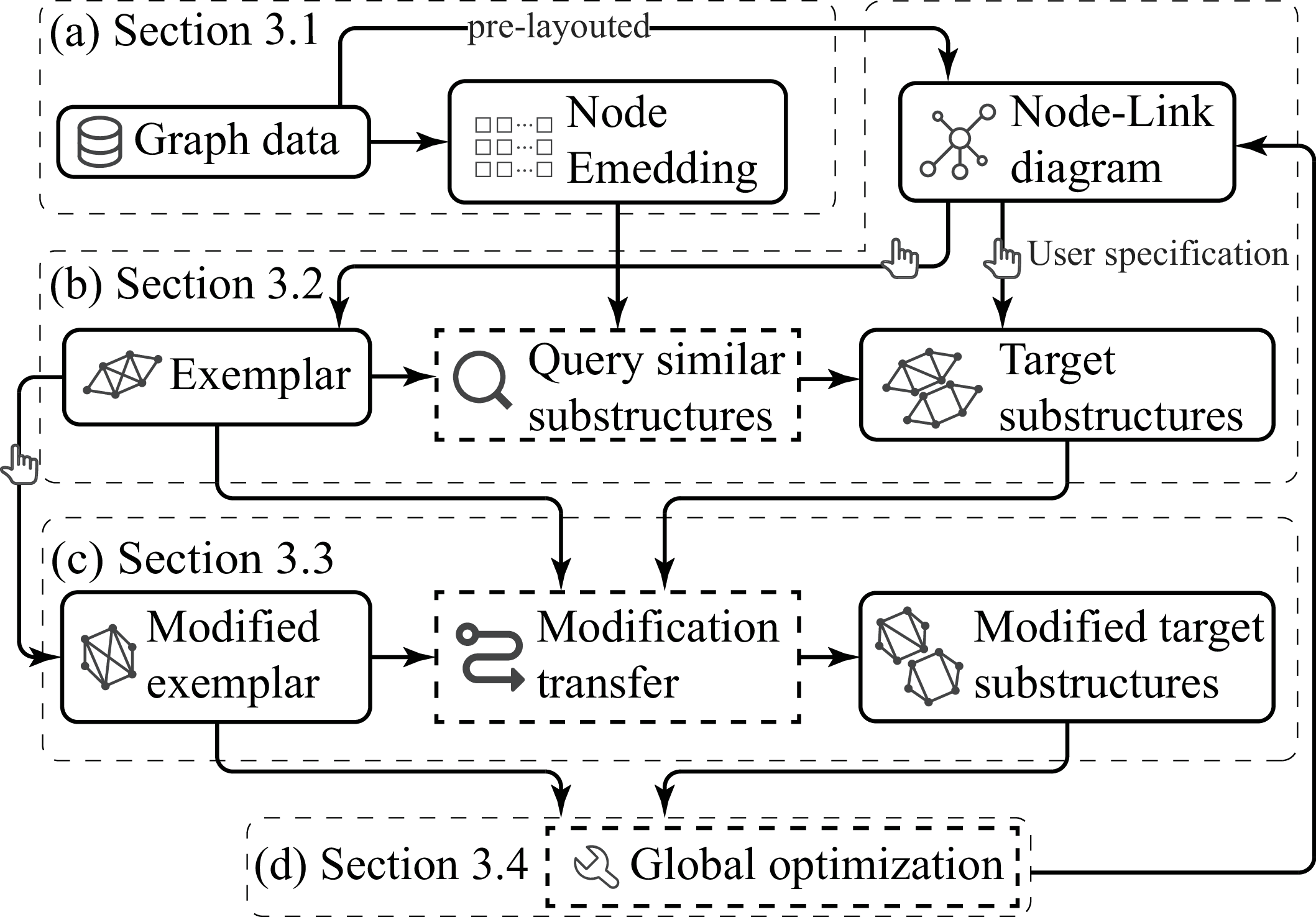}
    \caption{
    Workflow of our end-to-end system:
    (a) pre-computing node embeddings and laying-out
    a node-link diagram; (b) detecting similar target substructures with a specified exemplar; (c) transferring user modifications to similar substructures; (d) merging modified substructures into the entire layout with global optimization.}
    \label{fig:workflow}
\end{figure}

\replaced{We have designed and implemented an end-to-end tool for exemplar-based layout fine-tuning to reduce the manual workload of refining layout by suggesting fine-tuning candidates (similar substructures) and transferring user modifications to those candidates (Figure~\ref{fig:workflow}).}
{
An automatically generated node-link diagram may not be satisfying. For instance, a force-directed layout may lead to node overlapping or distorted cycle paths.
Fine-tuning can reduce visual clutter and improve visual aesthetics.
Transferring user modifications to other similar substructures in the entire graph can further gain performance improvement. After the exemplar is specified as the source substructure, 
substructures similar to the exemplar are retrieved in the underlying graph. 
The user modification on the exemplar is transferred onto similar substructures (the target substructures).
}
Our workflow has four steps:
\begin{compactenum}[\textbf{Step} 1.]
    \item
    Our algorithm calculates the node embedding of the entire graph to retrieve similar structures (Figure~\ref{fig:workflow}a). \added{A node-link diagram is generated with an initial layout of the entire graph.}
    \item
    The user specifies an exemplar from the entire graph.
    \replaced{
    Our similar structure-query technique using the method in~\cite{DBLP:journals/tvcg/ChenGHPNXZ19} retrieves several target substructures topologically similar to the exemplar (Figure~\ref{fig:workflow}b).}{Similar substructures are retrieved (Figure~\ref{fig:workflow}b).} The user can also specify target substructures from the node-link diagram.
    \item 
    \replaced{
    The user modifies the exemplar's layout. 
    Our modification transfer algorithm transfers the modifications to target substructures (Figure~\ref{fig:workflow}c).
    }{ Markers are generated by automatic graph matching methods or specified by the user on the exemplar and its counterparts, and then modifies the exemplar's layout. The modifications are transferred onto target substructures (Figure~\ref{fig:workflow}c).}
    \item 
    Our algorithm merges the modified substructures into the original
    graph through global optimization to smooth the boundaries. 
    The user can iterate from \textit{Step} $2$ to \textit{Step} $4$ to fine-tune the layout (Figure~\ref{fig:workflow}d).
\end{compactenum}

\subsection{Step 1. Node-embedding-based Representations}
Our approach uses a node-embedding technique to embed a node into a low-dimensional vector subject to its local topology. For a given exemplar, we employ the node-embedding-based representation to represent and retrieve similar substructures from the entire graph. In this way, we simplify the subgraph-retrieving problem to a similar multidimensional data-searching problem. 
Though various node-embedding representations~\cite{DBLP:conf/kdd/DonnatZHL18, DBLP:conf/kdd/GroverL16, DBLP:conf/cikm/HeimannSSK18, DBLP:journals/cn/MarcusS12, 10.1145/3097983.3098061} 
are compatible with our approach,
we leverage GraphWave~\cite{DBLP:conf/kdd/DonnatZHL18} \added[id=pan]{following the study conducted in~\cite{DBLP:journals/tvcg/ChenGHPNXZ19}.} \deleted[id=pan]{Section~\ref{sec:discussion} discusses the performance with different representations.}
\added[id=pan]{We pre-compute node embeddings because 
this process is
time-consuming.}

\subsection{Step 2. Specifying Exemplar and Targets} \label{sec:retrieval}
\begin{figure}[!tp]
    \centering
    \setlength{\belowcaptionskip}{-5pt}
    \includegraphics[width=1\columnwidth]{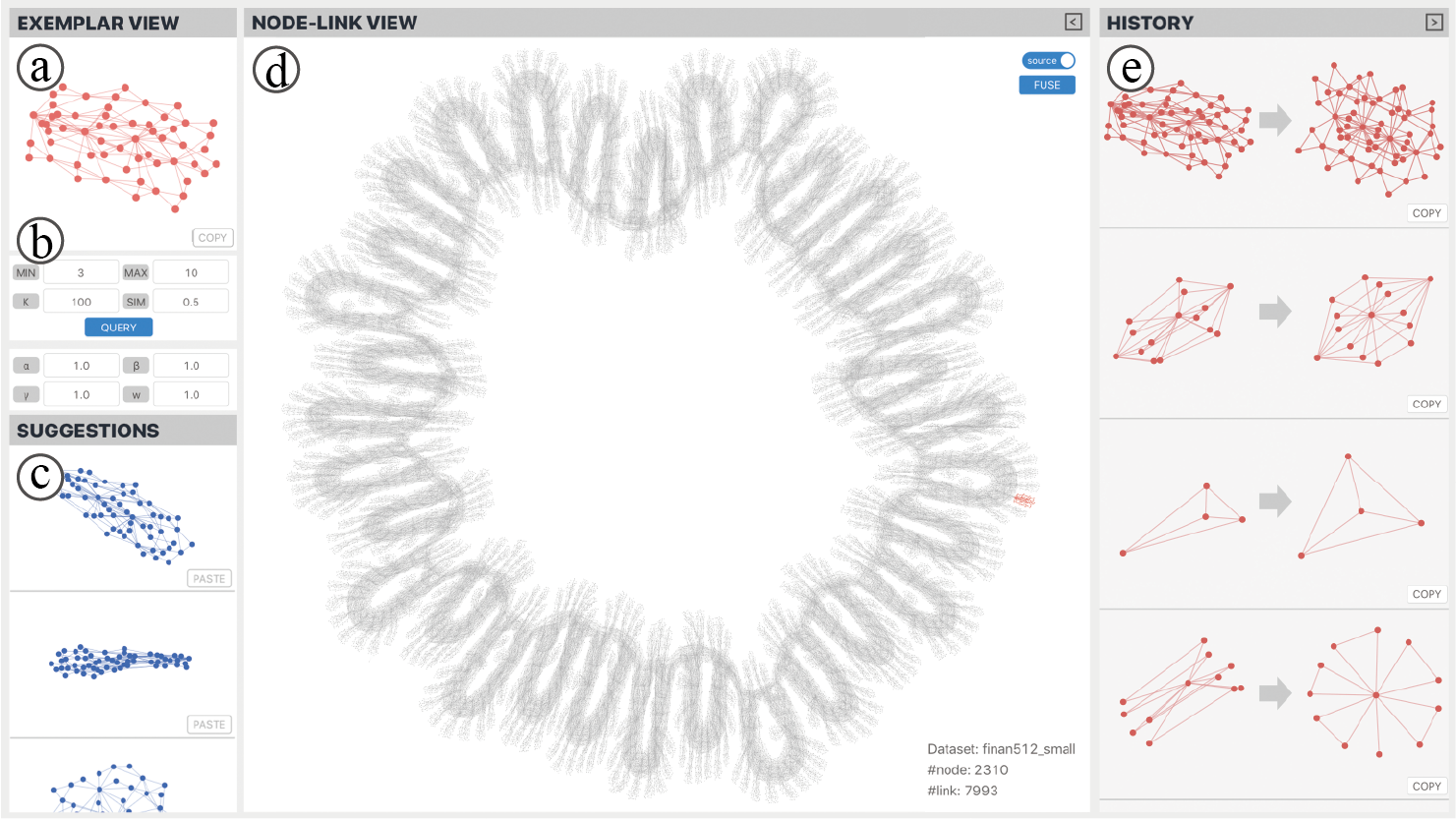}
    \caption{The user interface of our prototype system:
    (a) an exemplar view; (b) a control panel; (c) a suggestions gallery; (d) a node-link view; (e) a modification history view.}
    \label{fig:interface}
\end{figure}

The user can specify a substructure using the lasso interactions in the node-link diagram
\deleted[id=pan]{
After this,
candidate nodes that have similar embeddings with the exemplar's nodes are identified by the $k$-nearest neighbors algorithm. 
The connected substructures of candidate nodes thus induced are denoted as candidate substructures.
Similar substructures are further filtered by limiting their number of nodes and are ranked by Weisfeiler-Lehman similarity~\cite{DBLP:conf/nips/TogninalliGLRB19}.}
\added{(Figure~\ref{fig:interface}a).
We then use the similar structure-query technique 
in~\cite{DBLP:journals/tvcg/ChenGHPNXZ19} to retrieve a set of substructures that are potentially similar to the exemplar (Figure~\ref{fig:interface}c).}
Four parameters are used in the searching process.
The parameter $k$ is used in the $k$-nearest neighbors algorithm for retrieving similar nodes. 
\added[id=pan]{A large $k$ may introduce many candidate nodes in a huge connected substructure; it will be filtered out by parameter $max$. On the other hand, 
a small $k$ limits the number of candidate nodes, so that the probability of forming a connected substructure is small. The parameter $k$ should be tuned 
interactively.
}
We eliminate
similar substructures whose node number is less than the \textit{minimum count} ($min$) or more than the \textit{maximum count} ($max$).
\added[id=pan]{We suggest setting $min$ and $max$ 
to be 
close to the number of nodes in the exemplar (e.g., set $min$ to be half $\# nodes$ and $max$ to be twice $\# nodes$), so as to generate substructures of similar scale to the exemplar.} 
We also remove substructures whose Weisfeiler-Lehman similarity is less than the \textit{minimum similar threshold} ($\epsilon$).


Also, we let the user specify additional structures in the node-link diagram using lasso interactions. We regard both retrieved substructures and user-specified substructures as target substructures. 

\subsection{Step 3. User-driven Fine-tuning} \label{sec:tuning}
Our approach uses dragging interaction to interactively manipulate the exemplar's layout.
\added{The modification transfer algorithm described in Section~\ref{sec:ModificationTransfer} can transfer the exemplar's layout modifications to target substructures' layouts.
We design an interaction mode called ``format painter" (inspired by operations in Microsoft Word) to perform the modification transfer. 
After modifying the exemplar's layout, the user can transfer modifications to other target substructures using the ``copy'' and ``paste'' buttons. Our approach records modifications after the user clicks the ``copy'' button and transfers modifications into target substructures after the user clicks the ``paste'' button.}
\deleted{Then, we determine a set of markers to initialize the modification transfer. Many graph matching methods are represented for constructing nodes correspondences.
We test 6 graph matching methods to build correspondences: GA (graduated assignment)~\cite{gold1996graduated}, PM (probabilistic matching)~\cite{zass2008probabilistic}, SM (spectral matching)~\cite{leordeanu2005spectral}, SMAC (spectral matching with affine constraints)~\cite{cour2007balanced}, RRWM (re-weighted random walk matching)~\cite{cho2010reweighted} and FGMU (factorized graph matching for undirected graphs)~\cite{zhou2012factorized}. 
Three cases are performed with: three small collaboration networks taken from the Visualization-Publication dataset~\cite{Isenberg:2017:VMC}, three substructures taken from the Finan512 dataset~\cite{DBLP:journals/jgo/SoperWC04} (Figure~\ref{fig:correspondence}), and three circle structures taken from the Power-Network dataset~\cite{nr}. Our approach utilizes their results as markers in the modifications transferring. However, unpleasant results may be generated because these methods build correspondences for all nodes even if the nodes are not well matched. Our approach removes several poorly matched correspondences and selects others as markers by using the technique described in Algorithm~\ref{alg:corrselection}. In practice, FGMU is employed in our approach to generate markers. Note that, our approach also supports specifying markers manually to match user preferences.}

\deleted{The layout of the exemplar and target substructures are generated with the same algorithm before constructing correspondences, leading to more satisfying correspondences. In our implementation, FM$^3$~\cite{hachul2004drawing} is employed.}

\deleted{With specified markers, a ``format painter" interaction mode is designed to fulfill the modification transfer. The ``format painter" interaction is inspired by the ``Format Painter" in Microsoft Word:
after modifying the exemplar's layout, the ``format painter" can transfer modifications to other target substructures. 
New node positions of the target substructure are computed. After that, another substructure can be selected for ``painting".}

\subsection{Step 4. Global Layout Optimization}
\begin{figure}[!tp]
    \centering
    \setlength{\belowcaptionskip}{-5pt}
    \includegraphics[width=1\columnwidth]{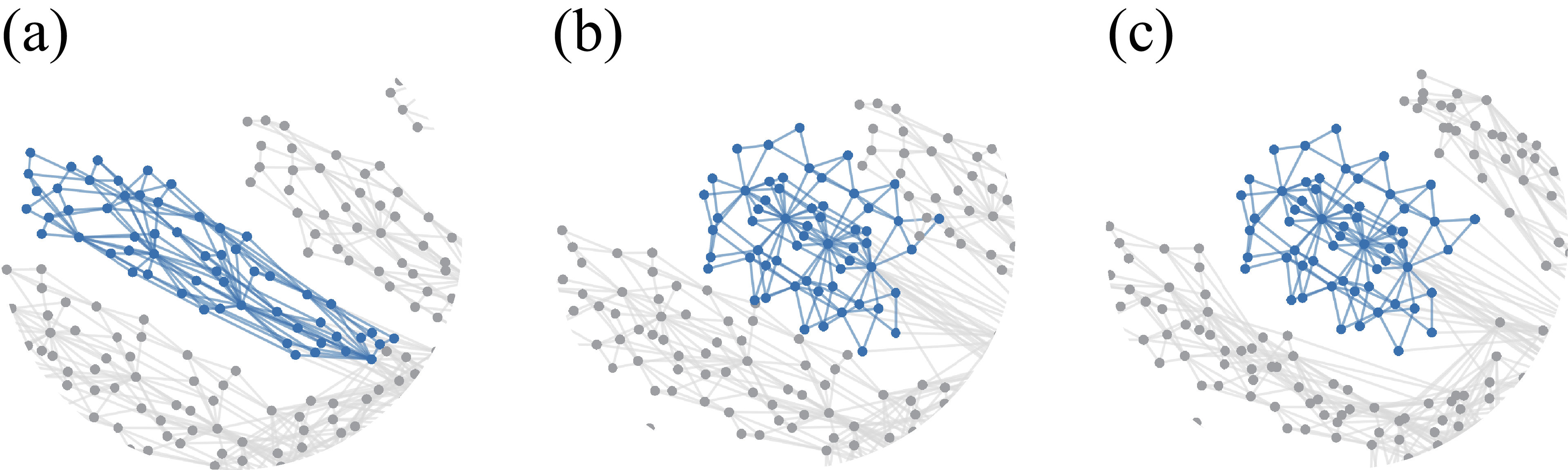}
    \caption{Global optimization in the Finan512 dataset case study; (a) original layout; (b) merging the modified target substructure into entire graph without any optimization; (c) merging modified target with our technique.}
    \label{fig:finan512detail}
\end{figure}

The exemplar and target substructures are parts of the entire graph. 
Directly merging 
the modified layout
into the entire graph may lead to abrupt boundaries of the modified substructures (Figure~\ref{fig:finan512detail}b). Thus, we perform a global optimization to preserve the smooth boundaries of the modified substructures (Figure~\ref{fig:finan512detail}c).
The optimization process is similar to the \textit{deforming} step,
like the stress-majorization layout~\cite{DBLP:conf/gd/GansnerKN04} (see Section~\ref{sec:layoutsimulation}).
We preserve details of the entire graph by minimizing the relative position displacements of each node pair. \deleted[id=pan]{Equation~\ref{eq:deforming} is solved to achieve a global smoothness.}
However, optimizing the entire graph is computationally expensive. 
We found that deforming the layout of the surroundings of the exemplar and target is to some extent adequate to reach smoothness. 
The surroundings of a structure are the induced subgraph of the entire graph whose nodes' distances are less than a given distance $d$ to the structure, \deleted[id=pan]{In practice, }
where $d$ is the maximum edge length in the entire graph.
\added[id=pan]{This ensures that nodes adjacent in both topology and Euclidean distance can be included.}

\subsection{Visual Interface}
We design and implement a visual interface, which consists of 4 parts:
\deleted{a) an \textit{exemplar view} supports exploring and modifying a specified exemplar;
b) a \textit{control panel} supports adjusting parameters;
c) a \textit{suggestion gallery} presents detected similar substructures and user-specified substructures;
d) a \textit{node-link diagram} shows the entire graph and lets the user explore and specify exemplars and targets;
e) a \textit{modification history view} records the change history.}
\textbf{The exemplar view}
(Figure~\ref{fig:interface}a) supports exploring and modifying a specified exemplar.
When the user finishes modifications,
the user can use ``format painter" to transfer modifications to the targets.
\added[id=pan]{\textbf{The control panel} (Figure~\ref{fig:interface}b) enables the user to adjust parameters of the modification transfer algorithm and the similar-substructure-retrieval algorithm.}
\textbf{The suggestion gallery}
(Figure~\ref{fig:interface}c) sequentially displays similar structures according to their Weisfeiler-Lehman similarities to the exemplar in node-link diagrams. 
In the meantime, the user can specify a structure in the node-link diagram as a suggestion; this is displayed at the top of the suggestion gallery. 
\textbf{The node-link view}
(Figure~\ref{fig:interface}d) provides visualizations with various graph-layout algorithms. The user can use the lasso to specify a substructure as an exemplar, which the exemplar view will then display.
\textbf{The modification history view}
(Figure~\ref{fig:interface}e) records layout change history applied to the exemplar. Each record relates to a piece of modification on the exemplar. The layouts before and after modifications are shown side by side. The user can reuse modifications in the history view for transferring. The most recent history is displayed at the top.

%% file: chapter/4.Algorithm.tex
\section{Modification Transfer in Graph Structures}
\label{sec:ModificationTransfer}

\begin{figure}[!t]
    \centering
    \includegraphics[width=\columnwidth]{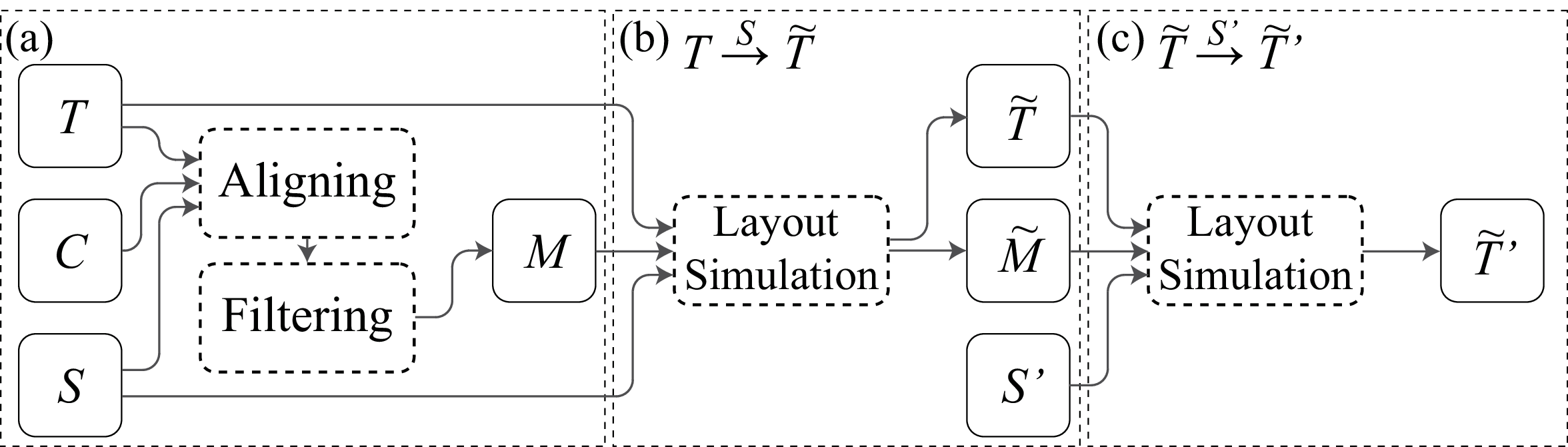}
    \caption{\added[id=pan]{Modification transfer. (a) Marker selection: aligning layouts of target $T$ and source $S$ first, and then selecting a set of markers $M$ from given correspondences $C$ between $S$ and $T$; (b) the first round of layout simulation: altering the layout $T$ to simulate $S$, which produces a new target structure layout $\tilde{T}$ and expands $M$ to $\tilde{M}$; (c) the second round of layout simulation: altering the layout $\tilde{T}$ to simulate $S^{\prime}$, which produces $\tilde{T}^{\prime}$.}}
    \label{fig:MT}
\end{figure}

Here we introduce a modification-transfer algorithm to transfer layout adjustments from one graph structure to another.
\added{We define terms in Table~\ref{tab:terms}.}
\replaced{Here, given a source graph structure layout $S=(V^s, E^s)$, user modifications change $S$ into a new layout $S^{\prime}$. And given a target graph structure layout $T=(V^t, E^t)$, we denote the modification transfer as a process of analogizing the modifications ($S \rightarrow S^{\prime}$) to the target graph ($T \rightarrow \tilde{T}^{\prime}$) in three steps:}
{A graph structure is denoted as $G=(V, E)$, where $V=\left\{v_1, v_2, \ldots, v_n\right\}, v_i \in \mathbb{R}^2$ is the positions of a set of $n$ nodes and $E=\left\{e_1, e_2, \ldots, e_m\right\}$ is a set of edges in $G$. Given a source structure $S=(V^s, E^s)$ (Figure~\ref{fig:layoutsimulation}a), user modifications change $S$ into a new layout $S^{\prime}$ (Figure~\ref{fig:layoutsimulation}b). Then, given a target structure $T=(V^t, E^t)$ (Figure~\ref{fig:layoutsimulation}d), we want to change $T=(V^t, E^t)$ with the same modifications. We denote modification transfer as the process of analogizing the modifications, which corresponds to applying the $S \rightarrow S^{\prime}$ modifications to $T$.}
\begin{compactenum}[\textbf{Step} 1]
    \item \added{\textbf{Marker selection}  first aligns $T$ and $S$ with correspondences $C$ generated by the graph-matching method and then selects some finely matched correspondences as markers (Figure~\ref{fig:MT}a).}
    \item \added{\textbf{Layout simulation ($T \stackrel{S}{\longrightarrow} \tilde{T}$)}
        alters the layout of the target from $T$ to $\tilde{T}$ to simulate $S$
        and expands $M$ to $\tilde{M}$ (Figure~\ref{fig:MT}b).}
    \item \added{\textbf{Layout simulation ($\tilde{T} \stackrel{S^{\prime}}{\longrightarrow} \tilde{T}^{\prime}$)}
        alters the layout of the target from $\tilde{T}$ to $\tilde{T}^{\prime}$ to simulate $S^{\prime}$ (Figure~\ref{fig:MT}c).}
\end{compactenum}
\added{Here, we perform two rounds of layout simulate because $S^{\prime}$ is usually different from $T$, directly deforming $T$ into the shape of $S^{\prime}$ can lead to unpleasing transfers.}


\begin{table}[!tp]
    \caption{\added[id=pan]{Definition of symbols. Here $G=(V, E)$ denotes a graph with its layout, where $V=\left\{v_1, v_2, \ldots, v_n\right\}, v_i \in \mathbb{R}^2$ contains the positions of a set of $n$ nodes and $E=\left\{e_1, e_2, \ldots, e_m\right\}$ is a set of $m$ edges in $G$.}}
    \centering
    \renewcommand\arraystretch{1.2}
    \begin{tabular}{l|l}
        \hline
        Symbol                         & Description                                                     \\
        \hline
        $S=(V^s, E^s)$                 & A source graph layout                                           \\
        $S^{\prime}$                   & A modified source graph layout                                  \\
        $T=(V^t, E^t)$                 & A target graph layout                                           \\
        $\tilde{T}=(\tilde{V}^t, E^t)$ & The target graph layout that simulates $S$'s layout             \\
        $\tilde{T}^{\prime}$           & The target graph layout that simulates $S^{\prime}$'s layout    \\
        $M$                            & The set of paired markers that matches $V^s$ to $V^t$           \\
        $(m^s_i, m^t_i) \in M$         & A pair of markers where $m^s_i \in V^s$ and $m^t_i \in V^t$     \\
        $C$                            & Correspondences between $S$ and $T$                             \\
        $(c^s_i, c^t_i) \in C$         & A correspondence pair where $c^s_i \in V^s$ and $c^t_i \in V^t$ \\
        $(v_i[x], v_i[y])$             & The $x$ and $y$ positions of node $i$                           \\
        $V(k)$                         & Positions of the node set $V$ in the iteration $k$              \\
        $\mathbf{R}$                   & A $3 \times 3$ affine transformation matrix                     \\
        \hline
    \end{tabular}
    \label{tab:terms}
\end{table}

\subsection {\added[id=pan]{Marker Selection}}

\added[id=pan]{The modification transfer algorithm relies on the correspondences between two structures, denoted as $C=\{(c^s_i, c^t_i)\}, 1 \leq i \leq \min(|V^s|, |V^t|)$.
    Any graph-matching method that produces injective correspondences is suitable for modification transfer.
    Six graph-matching methods~\cite{cho2010reweighted, cour2007balanced, gold1996graduated, leordeanu2005spectral, zass2008probabilistic, zhou2012factorized}
    are examined (see Suppl. Material\footnote{https://zjuvag.org/publications/exemplar-based-fine-tuning/} and Section~\ref{sec:quancomp} for comparison details). We employ Factorized Graph Matching (FGMU)~\cite{zhou2012factorized} because it achieves the best efficiency.}


Because graph-matching methods may depend on the graph layout, we layout the exemplar and target substructures with the same algorithm before constructing correspondences.
We employ FM$^3$~\cite{hachul2004drawing} \added{because it is one of the most efficient layout algorithms
    to our knowledge.}
The graph-matching methods can generate unpleasant
matching results because these methods build correspondences for all nodes even if they are not well matched.
\added{To examine their correspondences, we first align two graph structures $S$ and $T$
    according to the correspondences
    (described in Section~\ref{sec:layoutsimulation}).
    If the graph-matching method generates correct correspondences,
    we align two corresponding nodes
    together in the aligning step and almost all of their neighborhoods can possibly be matched. Thus, we implement
    the correspondences filtering algorithm (Algorithm~\ref{alg:corrselection}) to select ``fine'' correspondences ($(c^s_i, c^t_i)$) that satisfy:}
\begin{compactenum}[1)]
    \item \added[id=pan]{The distance between $c^s_i$ and $c^t_i$ is less than the average length of their adjacent edges multiplied by a given ratio ($r_d$); and}
    \item \added[id=pan]{ $c^s_i$'s neighbors are mostly matched to $c^t_i$'s neighbors (with a ratio greater than $r_u$).}
\end{compactenum}
\added[id=pan]{
    We fix $r_d$ and $r_u$ to be $2$ and $0.5$ in our implementation. A smaller $r_d$ and a larger $r_u$ lead to fewer, possibly more accurate correspondences. There is a trade-off between accuracy and number of correspondences.
    Here, we use these ``fine'' correspondences as a set of markers $M$ for the layout simulation.
    In addition,
    our approach also supports specifying markers manually to match user preferences.
    The user can click on two nodes, one in each of the exemplar and target substructures, to specify a pair of markers. }

\begin{algorithm}[!t]
    \renewcommand\arraystretch{1.2}
    \caption{ Correspondences filtering }
    \label{alg:corrselection}
    \begin{algorithmic}[1]
        \Require
        $S=(V^s, E^s)$: a source graph;
        $T=(V^t, E^t)$: a target graph;
        $C=\left\{(c^s_i, c^t_i)\right\}$: a set of correspondences;
        $r_u$: a minimum common neighbors ratio;
        $r_d$: a maximum distance ratio;
        \Ensure
        $M=\left\{(m^s_i, m^t_i)\right\}$: a set of markers;
        \State Init markers $M=\varnothing$ \label{code:initmarker}
        \For {each correspondence pair $(c^s_i, c^t_i)$}
        \State $ns \gets c^s_i$'s neighbors' corresponding nodes
        \State $nt \gets c^t_i$'s neighbors
        \State $nu \gets ns \bigcap nt$
        \If {$Count(nu) > Count(ns) \times r_u$ \textbf{or} $Count(nu) > Count(nt) \times r_u$}
        \State $ds \gets$ the mean length of adjacent edges of $c^s_i$
        \State $dt \gets$ the mean length of adjacent edges of $c^t_i$
        \State $d \gets$ distance between $c^s_i$ and $c^t_i$
        \If {$d < ds \times r_d$ \textbf{and} $d < dt \times r_d$}
        \State Add $(c^s_i, c^t_i)$ into $M$
        \EndIf
        \EndIf
        \EndFor
        \State \Return $M$;
    \end{algorithmic}
\end{algorithm}

\subsection {\added[id=pan]{Layout Simulation}} \label{sec:layoutsimulation}
\begin{figure}[t]
    \centering
    \includegraphics[width=1\columnwidth]{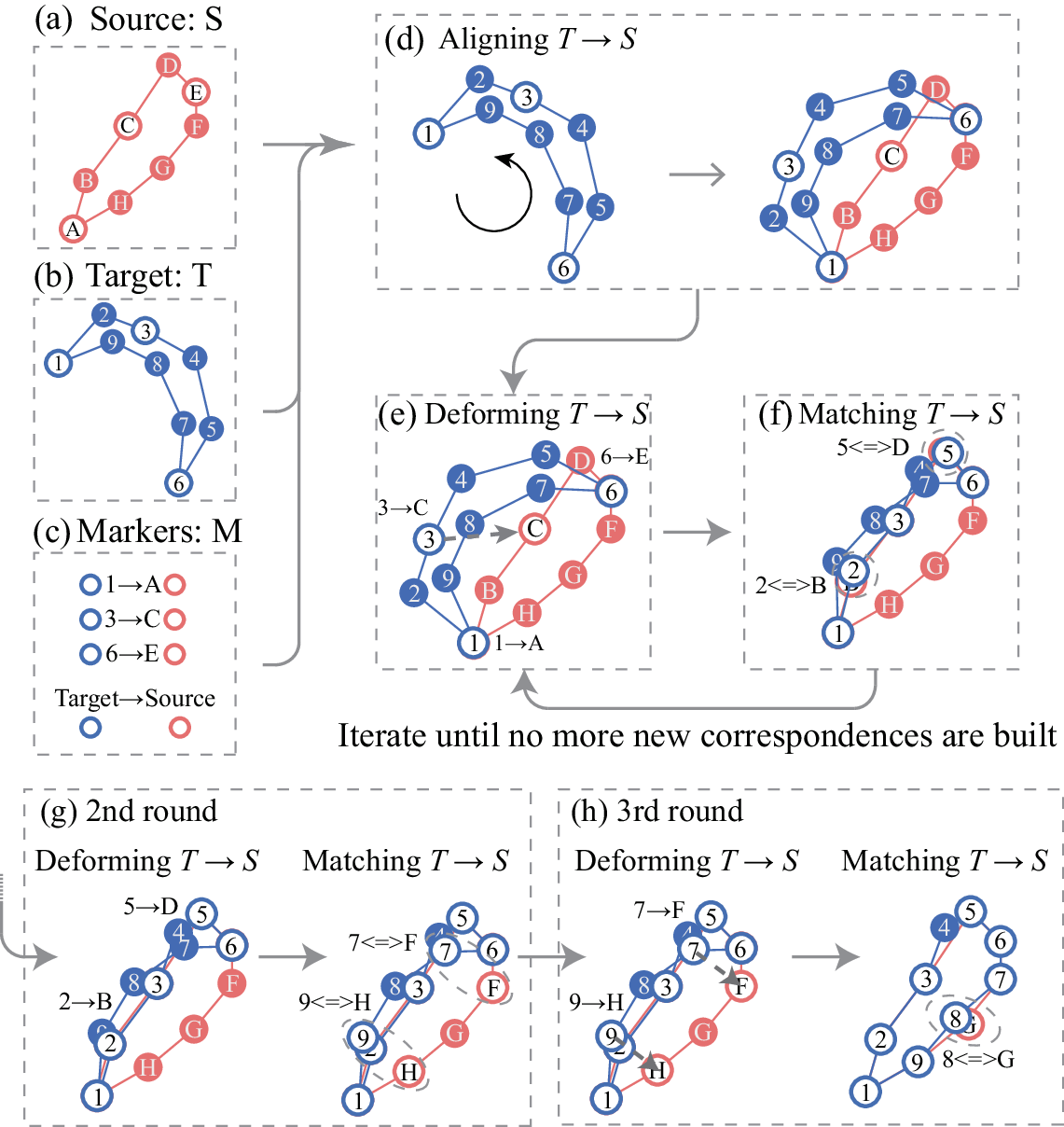}
    \setlength{\belowcaptionskip}{-10pt}
    \caption{\added[id=pan]{Layout simulation: altering the shape of a target structure $T$ to simulate the layout of a source structure $S$. (a) a source structure $S$; (b) a target structure $T$; (c) a set of markers $M$; (d) aligning $T$ to $S$ with markers $M$; (e) deforming $T$ into $S$ with markers $M$; (f) matching the nodes of $T$ to the nodes of $S$; two pairs of markers are constructed: \{(2,B) and (5,D)\}; (g) the second round of deforming and matching; two marker pairs are constructed: \{(7,F) and (9,H)\}; (h) the third round of deforming and matching, one pair of markers is constructed: \{(8,G)\}; Iterations are performed until no more new correspondences are built.}}
    \label{fig:layoutsimulation}
\end{figure}

\added[id=pan]{
    The goal of the layout simulation is to smoothly deform the markers of the target $T$ to those of the source $S$, while preserving the original layout of the target $T$ as much as possible (Figure~\ref{fig:layoutsimulation}).
}
We do this in three steps:
\textit{aligning}, \textit{deforming}, and \textit{matching}. The aligning step scales, rotates, and translates $T$ to minimize the dissimilarity to $S$ (Figure~\ref{fig:layoutsimulation}d). The deforming step alters the node positions of $T$ to simulate the shape of $S$ (Figure~\ref{fig:layoutsimulation}e). The matching step constructs correspondences between the nodes of $T$ and $S$ by searching their neighbors (Figure~\ref{fig:layoutsimulation}f).
These steps iteratively deform $T$ into the shape of \replaced[id=pan]{$S$}{$S^{\prime}$} until no more new correspondences are constructed.
\deleted[id=pan]{In practice, the process is performed to alter $T$ to simulate the shape of $S$ first  (Figures~\ref{fig:layoutsimulation} (e, f, and g)) and then to simulate $S^{\prime}$ (Figures~\ref{fig:layoutsimulation} (h, i, and j)). Directly deforming $T$ into the shape of $S^{\prime}$ can lead to unpleasing transferring (Figure~\ref{fig:layoutsimulation}k) because $S^{\prime}$ is usually quite different with $T$.}

\textbf{Aligning.}
We assume that both the topology and the
layout of the source structure
are similar to the target.
To minimize the layout difference, the node-link diagrams of $S$ and $T$ 
are
aligned.
\added[id=pan]{This step only transforms the global location and orientation of the target structure,
    not
    the 
    positions of individual nodes.
}

We use a small set of predefined markers to achieve an optimal alignment.
The markers are a set of paired nodes $M = \left\{(m^s_i, m^t_i)\right\}, m^s_i \in V^s, m^t_i \in V^t$ (Figure~\ref{fig:layoutsimulation}c). The markers on the source and the target are aligned by an affine transformation $\mathbf{R}$:
\begin{equation}
    \setlength{\abovedisplayskip}{5pt}
    \setlength{\abovedisplayshortskip}{5pt}
    \setlength{\belowdisplayshortskip}{5pt}
    \setlength{\belowdisplayskip}{5pt}
    \mathbf{R} = scale \times \left(\begin{array}{ccc}
            \cos \theta  & \sin \theta & tx \\
            -\sin \theta & \cos \theta & ty \\
            0            & 0           & 1
        \end{array}\right) \approx \left(\begin{array}{ccc}
            s  & h & tx \\
            -h & s & ty \\
            0  & 0 & 1
        \end{array}\right) \label{eq:transformation}
\end{equation}
where $scale$ is the scale coefficient, $\theta$ is the rotation angle, and $tx$ and $ty$ are the translation components. For the sake of simplicity, we use a linear approximation of $\mathbf{R}$ (after the approximately equal sign).
$\mathbf{R}$ is calculated by solving the minimization problem:
\begin{equation}
    \setlength{\abovedisplayskip}{5pt}
    \setlength{\abovedisplayshortskip}{5pt}
    \setlength{\belowdisplayshortskip}{5pt}
    \setlength{\belowdisplayskip}{5pt}
    \min_{\mathbf{R}} \sum_{i}^{|M|} ||\mathbf{R} m^{t}_i  - m^{s}_i ||^2,
\end{equation}
where $(m^t_i, m^s_i) \in M$ denotes one pair of markers.
The minimization problem is equivalent to the problem:
\begin{equation}
    \setlength{\abovedisplayskip}{5pt}
    \setlength{\abovedisplayshortskip}{5pt}
    \setlength{\belowdisplayshortskip}{5pt}
    \setlength{\belowdisplayskip}{5pt}
    \min_{\mathbf{T}} \sum_{M} ||\mathbf{A} (s, h, tx, ty)^T - \mathbf{b} ||^2,
\end{equation}
where $\mathbf{A}$ contains the positions of the markers in the target and $\mathbf{b}$ contains the positions of the markers in the source:
\begin{equation}
    \setlength{\abovedisplayskip}{5pt}
    \setlength{\abovedisplayshortskip}{5pt}
    \setlength{\belowdisplayshortskip}{5pt}
    \setlength{\belowdisplayskip}{5pt}
    \mathbf{A} = \left(\begin{array}{llll}
            {m^t_i}[x] & {m^t_i}[y]  & 1 & 0 \\
            {m^t_i}[y] & -{m^t_i}[x] & 0 & 1 \\
            \vdots
        \end{array}\right), \mathbf{b} = \left(\begin{array}{c}
            {m^s_i}[x] \\
            {m^s_i}[y] \\
            \vdots
        \end{array}\right), i=1,\ldots,|M|.
\end{equation}
${m^t_i}[x]$ and ${m^t_i}[y]$ are the positions of the target marker $m^t_i$ and ${m^s_i}[x]$ and ${m^s_i}[y]$ are the positions of the source marker $m^s_i$. The minimization problem can be solved by:
\begin{equation}
    \setlength{\abovedisplayskip}{5pt}
    \setlength{\abovedisplayshortskip}{5pt}
    \setlength{\belowdisplayshortskip}{5pt}
    \setlength{\belowdisplayskip}{5pt}
    (s, h, tx, ty)^T = \mathbf{A}^\dagger \mathbf{b} \label{eq:pseudoinverse},
\end{equation}
where $\mathbf{A}^\dagger$ is the Moore-Penrose pseudoinverse~\cite{moore1920reciprocal} of $\mathbf{A}$. Thus, the transformation can be defined as a linear function of the markers in the source.
With the affine transformation matrix $\mathbf{R}$, $T$ is transformed to align $S$ by a linear transformation (Figure~\ref{fig:layoutsimulation}d). \added[id=pan]{After modification transfer, the target layout is restored by an inverse process of the alignment step, so that it can be merged into the entire layout with the original rotation and scale.}

\textbf{Deforming.}
\deleted[id=pan]{
    Although the global location and orientation of the target structure is transformed after alignment, its shape (i.e., positions of individual nodes) is unchanged.} \added[id=pan]{The deforming step seeks to alter the shape of $T$ to simulate $S$. We design an energy function to represent the process:}
\begin{equation}
    \setlength{\abovedisplayskip}{5pt}
    \setlength{\abovedisplayshortskip}{5pt}
    \setlength{\belowdisplayshortskip}{5pt}
    \setlength{\belowdisplayskip}{5pt}
    \added[id=pan]{E = E_S + \gamma E_M},
\end{equation}
\replaced[id=pan]{where $\gamma$ is a weight parameter. The deforming step is equivalent to minimizing $E$. It seeks to force positions of target markers $m^t_i$ to approach source markers $m^s_i$ ($E_M$) while preserving the original layout information to reach a smooth deformation ($E_S$). Here, we denote $E_M$ as the sum of distances between pairs of markers:}{$T$ is deformed to simulate the shape of $S$ by forcing the target markers $m^t_i$ to approach $m^s_i$:}
\begin{equation}
    \setlength{\abovedisplayskip}{5pt}
    \setlength{\abovedisplayshortskip}{5pt}
    \setlength{\belowdisplayshortskip}{5pt}
    \setlength{\belowdisplayskip}{5pt}
    E_{M} = \sum_{i}^{|M|}||{m^t_i}-{m^s_i}||^{2}.
\end{equation}

\replaced[id=pan]{$E_S$ represents the layout change between $T$ and $\tilde{T}$, which is constructed by two items: }{To get a smooth deformation, relative positions of each node pair in the target should be preserved. Thus, other target nodes are also translated for achieving smoothness. We define the energy function of the smoothness as:}
\begin{equation}
    \setlength{\abovedisplayskip}{5pt}
    \setlength{\abovedisplayshortskip}{5pt}
    \setlength{\belowdisplayshortskip}{5pt}
    \setlength{\belowdisplayskip}{5pt}
    E_S = \alpha E_O + \beta E_D, \label{eq:smoothness}
\end{equation}
where $\alpha$ and $\beta$ are two weights, \replaced[id=pan]{$E_O$ is designed to preserve orientations of vectors between node pairs after the aligning step, and $E_D$ is designed to preserve distances between node pairs.}{and} $E_O$ is defined as:
\begin{equation}
    \setlength{\abovedisplayskip}{5pt}
    \setlength{\abovedisplayshortskip}{5pt}
    \setlength{\belowdisplayshortskip}{5pt}
    \setlength{\belowdisplayskip}{5pt}
    E_O = \sum_{i<j} w_{i j} || norm(v^t_i - v^t_j) - norm(\tilde{v}^t_i - \tilde{v}^t_j)||^2.
\end{equation}
Here, $norm(\cdot)$ denotes the normalization of a vector. $E_D$ is defined as:
\begin{equation}
    \setlength{\abovedisplayskip}{5pt}
    \setlength{\abovedisplayshortskip}{5pt}
    \setlength{\belowdisplayshortskip}{5pt}
    \setlength{\belowdisplayskip}{5pt}
    E_{D} = \sum_{i<j} w_{i j} (|| v^t_i - v^t_j || - || \tilde{v}^t_i - \tilde{v}^t_j ||)^2,
\end{equation}
where $w_{i j}$ is the weight related to the node pair $(v^t_i, v^t_j)$, and $(\tilde{v}^t_i, \tilde{v}^t_j)$ is a node pair of the target structure after deformation ($\tilde{T}$).
$w_{i j}$ is defined as:
\begin{equation}
    \setlength{\abovedisplayskip}{5pt}
    \setlength{\abovedisplayshortskip}{5pt}
    \setlength{\belowdisplayshortskip}{5pt}
    \setlength{\belowdisplayskip}{5pt}
    w_{i j}=\left\{\begin{array}{ll}
        w ||v^t_i - v^t_j||^{-2}, & \text { if } \left\{i, j\right\} \in E^t \\
        ||v^t_i - v^t_j||^{-2},   & \text { otherwise }
    \end{array}\right., \label{eq:weight}
\end{equation}
where $w$ is a preservation degree on the edges.
\added[id=pan]{
    Setting $w$ greater than $1$ makes the algorithm pay more attention to preserve orientations and length of edges.
}

\begin{figure}[t!p]
    \centering
    \setlength{\belowcaptionskip}{-5pt}
    \includegraphics[width=1\columnwidth]{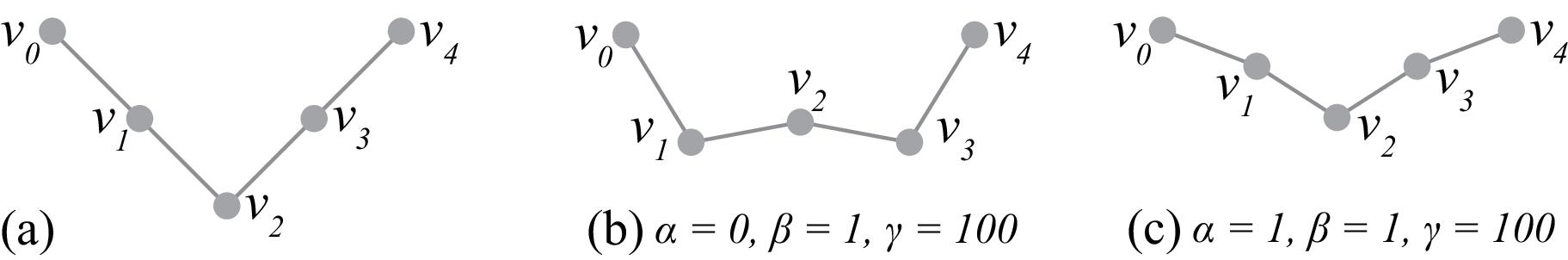}
    \caption{Different weighting schemes. The node $v_2$ is moved to a higher position while nodes $v_0$ and $v_4$ are fixed in their original positions. (a) is the original layout. (b) is the layout that preserves the distances with $\alpha=0, \beta=1$, and $\gamma=100$. (c) is the layout that keeps both orientations and distances with $\alpha=1, \beta=1$, and $\gamma=100$.}
    \label{fig:alphabeta}
\end{figure}

\deleted[id=pan]{$E_O$ and $E_D$ are designed to preserve the orientations and the lengths of vectors that link the node pairs in the target respectively.} Preferences on preservation of distances and orientations can be configured by balancing $\alpha$ and $\beta$. For example, when $\alpha$ is small, \replaced[id=pan]{the distances between node pairs}{the lengths of vectors} can be mostly preserved (Figure~\ref{fig:alphabeta}b). If we enlarge $\alpha$, the orientations can be better preserved (Figure~\ref{fig:alphabeta}c). Both weighting schemes are optional.
\added[id=pan]{The parameter $\gamma$ is used to configure the weight of moving marker positions in the target structure to their counterparts. A large $\gamma$ ensures that markers of $S$ and $T$ can be aligned.}

\deleted[id=pan]{The deforming step seeks to solve the minimization problem:
    \begin{equation}
        \setlength{\abovedisplayskip}{5pt}
        \setlength{\abovedisplayshortskip}{5pt}
        \setlength{\belowdisplayshortskip}{5pt}
        \setlength{\belowdisplayskip}{5pt}
        \min_{\tilde{v}^t} (E_S + \gamma E_M), \label{eq:deforming}
    \end{equation}
    where $\gamma$ is a weight. Setting $\gamma$ to be large adequately can fix the target markers' positions at their counterparts.}

\deleted[id=pan]{Optimizing Equation~\ref{eq:deforming} is identical to moving target markers' positions to their counterparts in the source while preserving the smoothness.} Following the optimization process in the stress majorization technique~\cite{DBLP:conf/gd/GansnerKN04}, $E_O$ and $E_D$ can be minimized by iteratively solving:
\begin{equation}
    \setlength{\abovedisplayskip}{5pt}
    \setlength{\abovedisplayshortskip}{5pt}
    \setlength{\belowdisplayshortskip}{5pt}
    \setlength{\belowdisplayskip}{5pt}
    L_{w}^{V^t(k)} V^t(k+1) = L_{w}^{V^t} V^t,
\end{equation}
and
\begin{equation}
    \setlength{\abovedisplayskip}{5pt}
    \setlength{\abovedisplayshortskip}{5pt}
    \setlength{\belowdisplayshortskip}{5pt}
    \setlength{\belowdisplayskip}{5pt}
    L_{w} V^t(k+1) = L_{w}^{V^t(k)} V^t(k),
\end{equation}
where $V^{t}(k)$ and $V^{t}(k+1)$ are the target nodes in time $k$ and $k + 1$. $L_w$ and $L_{w}^{V^t}$ are two weighted Laplacian matrices defined as:
\begin{equation}
    \setlength{\abovedisplayskip}{5pt}
    \setlength{\abovedisplayshortskip}{5pt}
    \setlength{\belowdisplayshortskip}{5pt}
    \setlength{\belowdisplayskip}{5pt}
    \begin{aligned}
        (L_{w}^{V^t})_{i j} & = \left\{\begin{array}{ll}
            - w_{i j} inv(||V^t_i - V^t_j||),    & i \neq j \\
            \sum_{l \neq i} (L_{w}^{V^t})_{i l}, & i = j    \\
        \end{array}\right.  \\
        (L_{w})_{i j}       & = \left\{\begin{array}{ll}
            - w_{i j},               & i \neq j \\
            \sum_{l \neq i} w_{i l}, & i = j    \\
        \end{array}\right.,
    \end{aligned}
\end{equation}
and the definition of $L_{w}^{V^t(k)}$ is similar to $L_{w}^{V^t}$ except that $v^t_i$ and $v^t_j$ are replaced by their counterparts in time $k$. The process is repeated until the target layout stabilizes.

\textbf{Matching.}
The matching step constructs node correspondences between $S$ and $T$.
\replaced[id=pan]{Any node pair $(v^s_i, \tilde{v}^t_j), i \leq |V^s|, j \leq |\tilde{V}^t|$ that satisfies $|| v^s_i - \tilde{v}^t_j || < r_j$ is identified as one candidate correspondence.}{
    For each node $\tilde{v}^t_i$ in $\tilde{T}$, nodes in the source subject to $||v^s_j - \tilde{v}^t_i|| < r_i$ are identified as its candidate nodes (Figure~\ref{fig:layoutsimulation}f).}  \replaced[id=pan]{We consider that $r_j$ should be adaptive to different $\tilde{v}^t_j$, and thus, we associate $r_j$ to the mean length of $\tilde{v}^t_j$'s adjacent edges. By default}{In practice}, $r_j$ is set to be twice the mean length of the adjacent edges \added[id=pan]{to avoid filtering out too many candidate node pairs}.
To avoid overlapping, \replaced[id=pan]{correspondences should be injective.}{only one corresponding node is identified for a source node $v^s_j$.} This maximum assignment problem can be solved by the Hungarian algorithm~\cite{kuhn1955hungarian, kuhn1956variants}. \added[id=pan]{Here, we use distances between node pairs as the cost in the Hungarian algorithm.}


Adequate correspondences can yield accurate modification transfer.
Thus, the \textit{aligning}, \textit{deforming}, and \textit{matching} steps are iteratively performed by using the already-built correspondences or markers.
For example, Figures~\ref{fig:layoutsimulation}(e-f) show the first round of deformation. With three markers, the target can not faithfully mimic the shape of the source. Additional correspondences are constructed by searching neighbors (Figure~\ref{fig:layoutsimulation}f). Two more \textit{deforming} and \textit{matching} rounds improve the accuracy (Figures~\ref{fig:layoutsimulation}(g-h)). The iteration stops until the number of correspondences
no longer increases.
$\tilde{T}$ is often similar to $S$ after deformation (Figure~\ref{fig:layoutsimulation}h). After that, layout simulation is performed again to alter the deformed target $\tilde{T}$ into the modified source $S^{\prime}$ (Figure~\ref{fig:MT}c).


%% file: chapter/5.Experiments.tex
\section{Results and Evaluation}

We implement our system in a browser-based architecture.
The front-end application is developed with JavaScript using React and D3. The back-end server uses Python 3.7.5 with flask, networkx, numpy, and scipy. All experiments are performed on a Macbook Pro laptop with an Intel Core i7-7820HQ CPU (2.9 GHz) and 16 GiB RAM. 

\begin{figure}[t!p]
    \centering
    \setlength{\belowcaptionskip}{-5pt}
    \includegraphics[width=1\columnwidth]{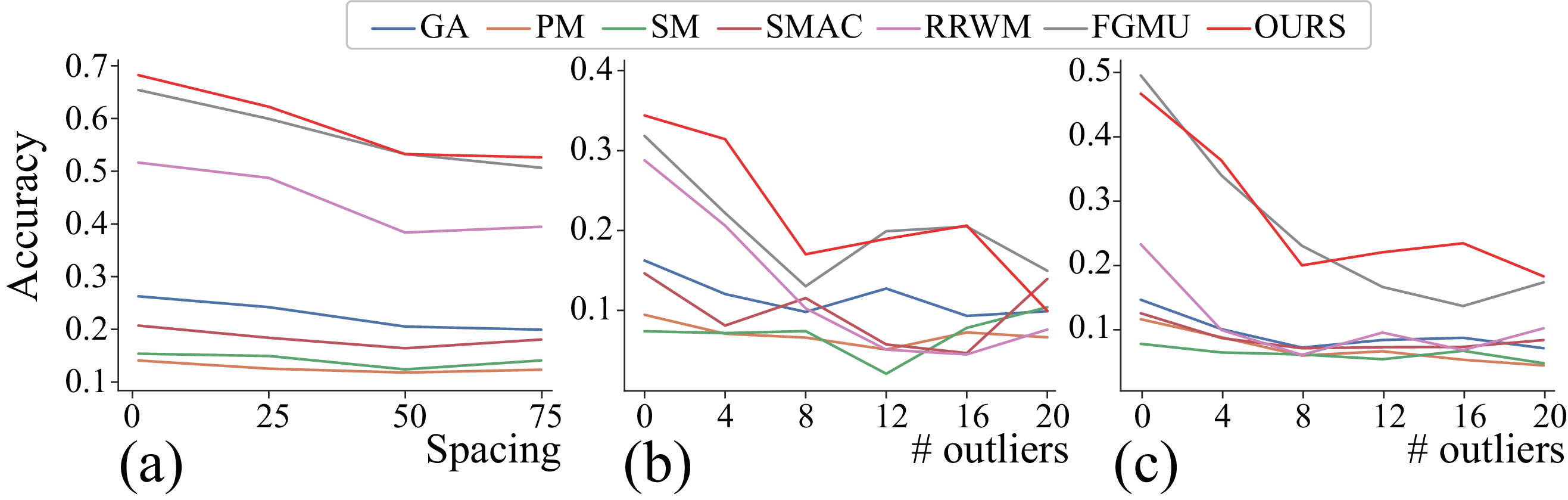}
    \caption{Quantitative comparison of several conventional graph matching methods and our approach: (a) average accuracy of different frame spacing in the CMU-house-image dataset; (b) average accuracy of different numbers of outliers in the Motorbike-image dataset; (c) average accuracy of different numbers of outliers in the Car-image dataset.}
    \label{fig:quancomp}
\end{figure}

\begin{figure*}[htb]
    \centering
    \setlength{\belowcaptionskip}{-5pt}
    \includegraphics[width=2\columnwidth]{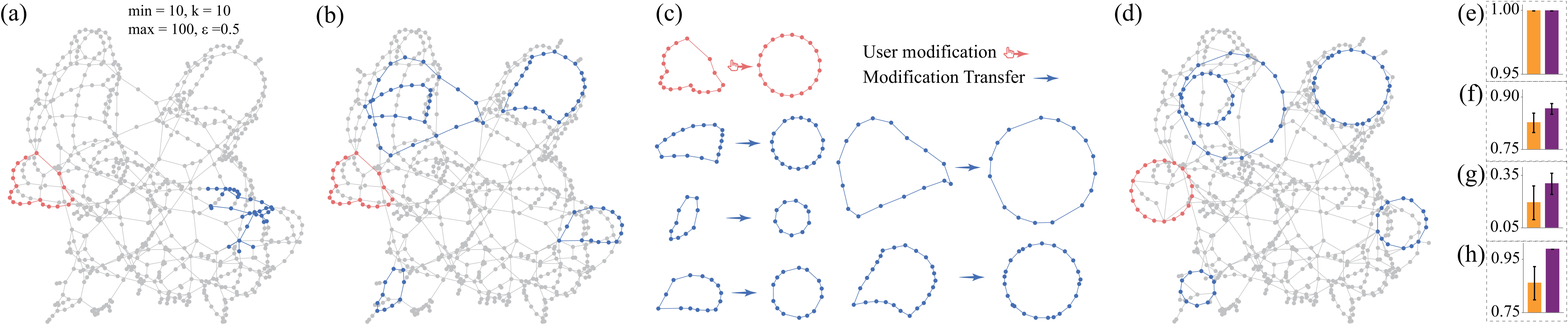}
    \caption{Case study with Power-Network dataset~\cite{nr}: (a) \added{an exemplar (in red) and two retrieved substructures (in blue, which are overlapped) overlaid on a network depicted using FM$^3$~\cite{hachul2004drawing}.} \added[id=pan]{(b) Two retrieved substructures are discarded. And several target substructures are specified manually.}(c) \replaced[id=pan]{The shape of the exemplar}{An exemplar is specified and its shape} is changed to a circle. \deleted[id=pan]{Several target substructures and their markers are specified.}  Modification transfer alters the node positions of targets to simulate the exemplar's shape. (d) All modified substructures are merged into the graph by an automatic optimization. \added{(e-h) Readability before (orange) and after (purple) modification transfer measured by four readability criteria (from top to bottom: crosslessness, minimum-angle metric, edge-length variant, and shape-based metric);; error bars depict 95\% confidence intervals.}}
    \label{fig:powernetwork}
\end{figure*}

\subsection {Quantitative Comparison} \label{sec:quancomp}
Our approach 
uses
a set of markers generated by graph-matching methods for modification transfer. 
We compared conventional graph-matching methods to ours
using the following benchmark datasets \added[id=pan]{with manually labeled ground truth}:

\begin{compactenum}[\bfseries 1)]

\item \textbf{The CMU-house-image dataset}~\cite{zhou2012factorized} 
contains 111 frames of a house with 30 landmarks. We randomly remove 5 landmarks and generate a graph with Delaunay triangulation that connects landmarks for each frame. Frames are paired spaced by 0, 25, 50, and 75 frames, yielding 444 pairs.

\item \textbf{The Car-and-Motorbike image dataset}~\cite{DBLP:journals/ijcv/LeordeanuSH12} has 30 pairs of car images and 20 pairs of motorbike images. 
We used Delaunay triangulation to generate graphs for each image, added 0, 4, 8, 12, 16, and 20 outliers randomly, and removed unconnected nodes, yielding 222 pairs of graphs.

\end{compactenum}

Node-link diagrams of these datasets are generated by the well-studied force-directed layout algorithm. We compare the accuracy of graph matching results. Figure~\ref{fig:quancomp} shows the average matching accuracy on different datasets. Our approach works slightly better than FGMU and exceeds other methods, meaning that our improvements on FGMU can generate more accurate results in most cases. \added[id=pan]{Note that graphs in these benchmark datasets are smaller than those in the case studies.}

\subsection{Case Studies}
We show how our exemplar-based layout fine-tuning approach works in three case studies.

We used FM$^3$~\cite{hachul2004drawing} to generate the layout of the \textbf{Finan512 dataset} from the University of Florida Sparse Matrix Collection~\cite{10.1145/2049662.2049663}, which is generated from multistage stochastic financial modeling~\cite{DBLP:journals/jgo/SoperWC04}. 
The graph consisted of 74,752 nodes and 261,120 edges rendered in WebGL 
(Figure~\ref{fig:Finan512}a). 
We saw several ``donut-like" substructures.

Next, we specified a substructure (here called an \textit{exemplar}) for fine-tuning (Figure~\ref{fig:Finan512}b). 
We retrieved similar substructures 
using
$k = 200$, $min=10$, $max=100$, and $\epsilon=0.95$ (Figure~\ref{fig:Finan512}a). 
\deleted{We find that these substructures are potentially isomorphic (though because their overall layout quality is poor, it is hard to identify whether they are isomorphic or similar).} To verify the topology of these substructures, we select 
target substructures as the five most similar and five most dissimilar substructures according to their Weisfeiler-Lehman similarities to the exemplar (Figure~\ref{fig:Finan512}c).
In addition, to fine-tune the ``donut" subgraph, we use substructures around it 
as target substructures.

We interactively modified the exemplar into a layout with a distinguishable structure (Figure~\ref{fig:Finan512}e).
After modification transfer, these substructures became clearer (Figures~\ref{fig:Finan512}(f, g)). 
\deleted{Most of them are isomorphic. Nodes in several non-isomorphic but similar structures are placed in harmony with the exemplar. 
}

Our smooth merging scheme generated 
visually pleasing details compared to direct merging without any optimization. For example, the 
boundary
of the substructure in Figure~\ref{fig:finan512detail}c is easier to distinguish than the one without optimization in Figure~\ref{fig:finan512detail}b.

\textbf{The Power-Network dataset} is collected from the Network Data Repository~\cite{nr}, which abstracts a power system: the nodes encode buses and edges are the transmission lines among the nodes. The network contains 662 nodes and 906 edges. 
A multilevel graph layout implemented by Tulip~\cite{auber2004tulip} and \added[id=pan]{OGDF~\cite{DBLP:reference/crc/ChimaniGJKKM13}} is employed to layout the network (Figure~\ref{fig:powernetwork}a).

To reveal transmissions among a set of buses that may form a cycle\deleted[id=pan]{, nodes are formed into a circle shape.}, we specify a set of nodes as an exemplar (Figure~\ref{fig:powernetwork}a, in red).\deleted[id=pan]{The exemplar is interactively modified into a circle.} \replaced[id=pan]{With $min=10$, $max=100$, $k=5$, and $\epsilon=0.5$, two overlapped structures are retrieved (Figure~\ref{fig:powernetwork}a, in blue). The retrieved structures are not topologically similar to the exemplar, because our technique detects embedding-similar structures, which are potentially similar to the exemplar.
}{The retrieved similar structures are not satisfactory.}
Thus we explore the node-link diagram to specify target substructures. Several sets of nodes that may form cycles are specified as target substructures (Figure~\ref{fig:powernetwork}b, in blue). Connections among these nodes are obscured by the visual clutter. \added[id=pan]{The exemplar is interactively modified into a circle.} Each target is deformed into a circle-like shape by transferring modifications (Figure~\ref{fig:powernetwork}c). Now the connections among nodes are far more distinguishable (Figure~\ref{fig:powernetwork}d) than the original layout.

\begin{figure}[!tp]
    \centering
    \setlength{\belowcaptionskip}{-5pt}
    \includegraphics[width=1\columnwidth]{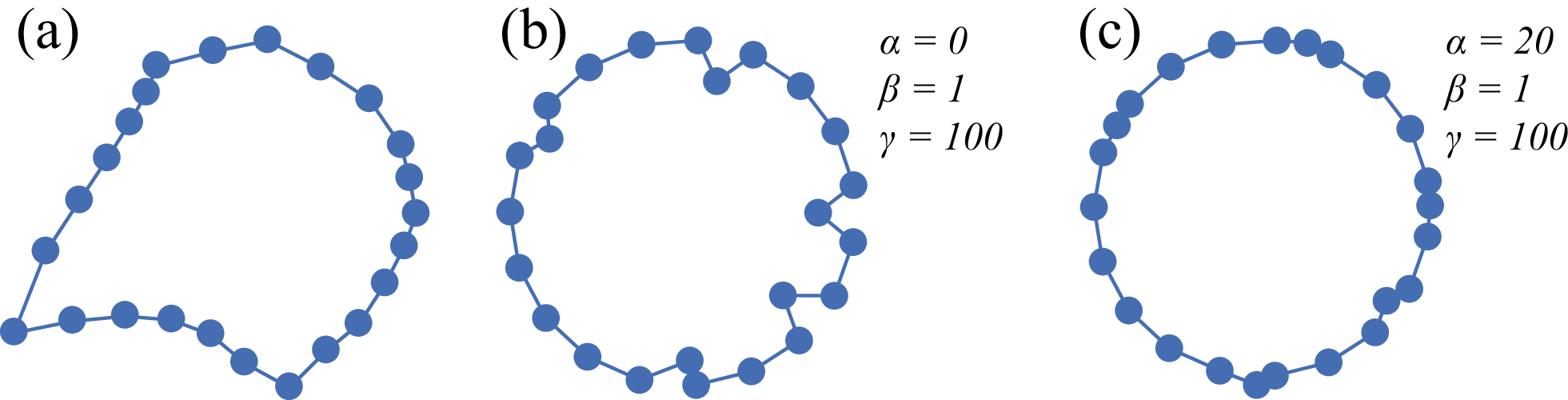}
    \caption{Different weighting schemes for the Power-Network dataset. (a) a target substructure; (b) a low preservation on orientations with $\alpha = 0$, $\beta = 1$, and $\gamma = 100$; (c) a large preservation on orientations with $\alpha = 20$, $\beta = 1$, and $\gamma = 100$.}
    \label{fig:pownetweight}
\end{figure}

We increase $\alpha$ to increase the degree of orientation preservation, which means that orientations of edges tend to remain unchanged. This makes the shape of the modified target substructure smoother (Figure~\ref{fig:pownetweight}c).
Because the edge lengths before modification transfer are not identical (Figure~\ref{fig:pownetweight}a), solely preserving distances can lead to unsatisfying deformations. For example, setting $\alpha$ in Equation~\ref{eq:smoothness} to be zero generates irregular polygons (Figure~\ref{fig:pownetweight}b).

\begin{figure*}[t!hbp]
    \centering
    \setlength{\belowcaptionskip}{-5pt}
    \includegraphics[width=2\columnwidth]{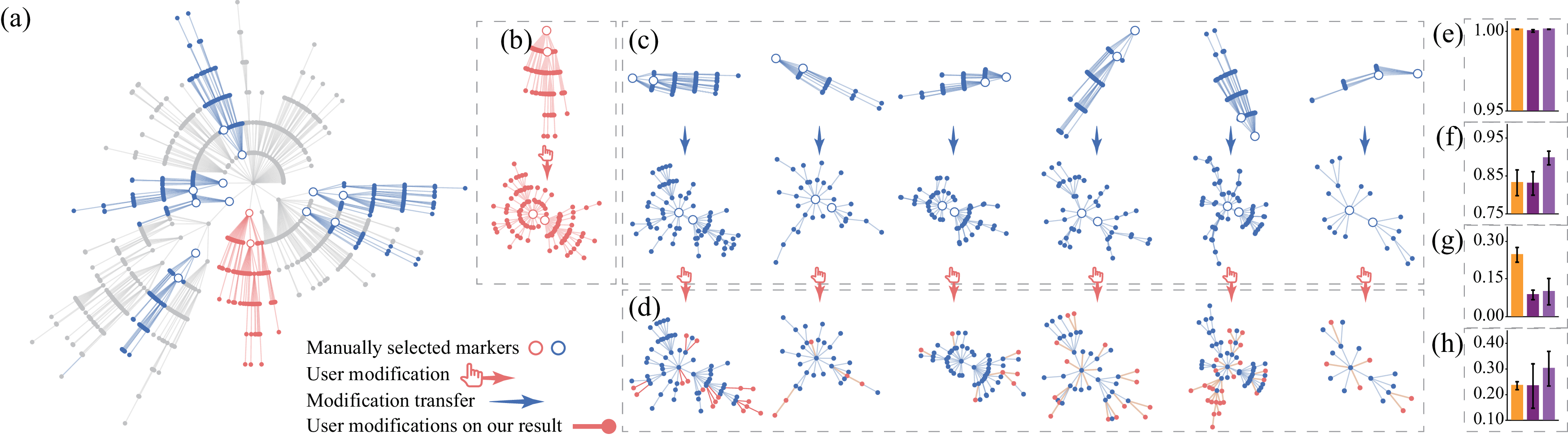}
    \caption{The Price\textunderscore1000 dataset~\cite{DBLP:journals/cgf/KruigerRMKKT17}. (a) A radial tree layout. (b) A specified exemplar in which we specify two nodes with the two largest degrees as markers. This is modified into a radial tree layout interactively. (c) Targets and their counterparts after modification transfer. (d) Modified targets after several slight user modifications \added[id=pan]{(in red)}. \added{(e-h) Readability before (orange) and after (purple) modification transfer measured by four readability criteria (from top to bottom: crosslessness, minimum-angle metric, edge-length variant, and shape-based metric); error bars depict 95\% confidence intervals..}}
    \label{fig:price1000}
\end{figure*}

\textbf{The Price\textunderscore1000 dataset} is a tree\deleted[id=pan]{-like graph} from tsNET~\cite{DBLP:journals/cgf/KruigerRMKKT17} that consists of 1000 nodes and 999 edges. We layout the graph with a simple radial tree layout algorithm~\cite{DBLP:conf/infovis/Jankun-KellyM03} (Figure~\ref{fig:price1000}a), and find that sibling nodes are overlapped due to the space constraint. 

We select one representative subtree as an exemplar. To reduce visual clutter,this is reconfigured into a radial tree layout (Figure~\ref{fig:price1000}b). To reconfigure other interested subtrees, we specify two nodes of the exemplar as markers, and the algorithm transfers modifications on the exemplar to other subtrees (Figure~\ref{fig:price1000}c).

Although there are some unpleasing details, their layouts are similar to the exemplar's. Rather than interactively reconfiguring these subtrees from the original layout, our approach requires only a few slight modifications \added[id=pan]{according to the minimum angle and the symmetry aesthetic metrics~\cite{DBLP:journals/vlc/Purchase02}} to tune the details (Figure~\ref{fig:price1000}d) because it generates an initial layout for each subtree. 

\added[id=pan]{
\textbf{Readability.} To evaluate the readability of the results generated by our approach, we use the measurements (crosslessness, minimum-angle metric, edge-length variation, and shape-based metric) in~\cite{DBLP:journals/tvcg/KwonCM18} to test readability improvement. All these measurements are normalized. Larger values of the measurements suggest higher readability except edge-length variation. Results of readability measurements for the Finan512 dataset, the Power-Network dataset and the Price\textunderscore1000 dataset are given in Figures~\ref{fig:Finan512}(h,i,j,k), Figures~\ref{fig:powernetwork}(e,f,g,h), Figures~\ref{fig:price1000}(e,f,g,h), accordingly. Bars representing measurement values before modification transfer are in orange and bars after modification transfer are in purple.
Note that, in the case study with the Price\textunderscore1000 dataset, we also measure the readability after slight modifications (in light purple). Results show that our approach improves readability in most cases.
}

\subsection{User Study} \label{sec:userstudy}
We conducted a \replaced[id=pan]{within-participant}{between-subject} experiment in which we asked participants to fine-tune structures’ layouts in three modes:

\added{\textbf{1) Baseline manual}}: \deleted[id=pan]{manually} mouse dragging without our approach\deleted{(manual, $T_m$)};

\added{\textbf{2) Our semi-automatic method}}\deleted{: our approach} with markers specified by user\deleted{(semi-automatic, $T_s$)};

\added{\textbf{3) Our fully automatic method}} \deleted{: our approach} with markers initialized by filtering the results of FGMU\deleted{(fully automatic, $T_f$)}.

\textbf{\added[id=pan]{Task.}} Participants performed a task involving modifying the structure on the screen \replaced[id=pan]{according to the expert's modifications on the exemplar}{to the target structure made by visualization expert}. \added{Twenty substructures from four real-world datasets are used.}

\begin{figure}[!tp]
    \centering
    \setlength{\belowcaptionskip}{-5pt}
    \includegraphics[width=1\columnwidth]{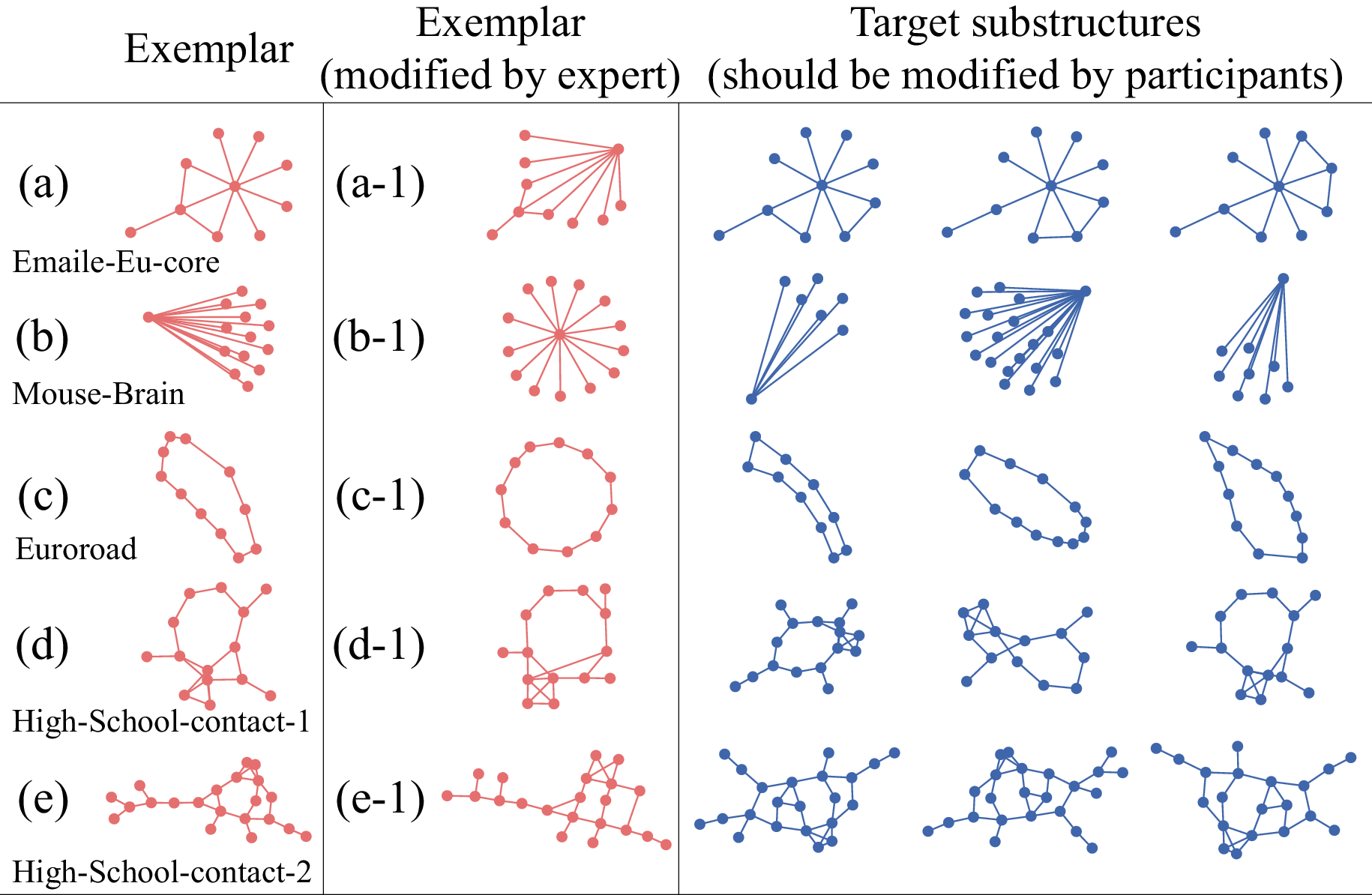}
    \caption{\added[id=pan]{Data samples in the user study.} Five exemplars and 15 target substructures were extracted from the four datasets. (a) \added[id=pan]{is} extracted from the Email-Eu-core dataset~\cite{paranjape2017motifs}; 
    (b) \added[id=pan]{is from} the Mouse-Brain dataset~\cite{DBLP:conf/miccai/FournierLE16}; 
    (c) \replaced[id=pan]{is from the Euroroad dataset~\cite{vsubelj2011robust};}{ $D_{HS1}$ and (d) $D_{HS2}$ generated from the High-School-contact dataset;}
    \replaced[id=pan]{(d) and (e) are from the High-School-contact dataset~\cite{10.1371/journal.pone.0136497}.}{(e) $D_R$ taken from the Euroroad dataset; (f-j) exemplars reconfigured by an expert majored in graph visualization.}}
    \label{fig:usdata}
\end{figure}

\textbf{Datasets.} \replaced[id=pan]{A graph visualization expert helped us define the 20 total substructures used in the study. He first chose five exemplar substructures from four real-world datasets and then specified three target structures for each of the five exemplars (Figure~\ref{fig:usdata}). Graphs generated from four real-world datasets have already been laid out with FM$^3$~\cite{hachul2004drawing}. Substructures are extracted with node positions. The expert was also asked to modify five exemplars' layouts to support our task.
}
{Four datasets were used in this study.}

\textbf{1)} The Email-Eu-core dataset~\cite{paranjape2017motifs} is a time-varying email contact network in a large European research institution with 986 nodes and 332,334 contacts. Email communications within every 24 hours form a graph, yielding a total of 803 snapshots with 855 connected subgraphs. \added[id=pan]{We obtained the first exemplar and its three target structures from Email-Eu-core dataset (Figure~\ref{fig:usdata}a). The expert modified the exemplar into a fan-like shape (Figure~\ref{fig:usdata}a-1).}

\textbf{2)} The Mouse-Brain dataset~\cite{DBLP:conf/miccai/FournierLE16} consists of 986 nodes and 1,536 edges. Nodes represent the mouse visual cortical neurons and edges are fiber tracts connecting one neuron to another. \added[id=pan]{We obtained the second exemplar and its three target structures from the Mouse-Brain dataset (Figure~\ref{fig:usdata}b). The expert modified the exemplar into a star-like shape in which the interior node stays in the center and the leaves are placed evenly around the interior (Figure~\ref{fig:usdata}b-1).}

\textbf{3)} The Euroroad dataset~\cite{vsubelj2011robust} is a road network mostly in Europe. Nodes represent cities and an edge between two nodes denote that they are connected. The network consists of 1,174 nodes and 1,417 edges. \added[id=pan]{We obtained the third exemplar and its three target structures (Figure~\ref{fig:usdata}c) are extracted from the Euroroad dataset~\cite{vsubelj2011robust}. The expert modified the exemplar into a round circle (Figure~\ref{fig:usdata}c-1).}

\textbf{4)} The High-School-contact dataset collected from the SocioPatterns initiative~\cite{10.1371/journal.pone.0136497} consists of 180 nodes and 45,047 contacts. We created a temporal network following the procedure in~\cite{von2009system}. 
\added[id=pan]{The last two exemplars and six target structures were obtained from the High-School-contact dataset (Figures~\ref{fig:usdata}(d, e)). The expert modified one exemplar into a shape in which the inner circle is laid out as a regular polygon and the surrounding nodes are placed orthogonally (Figure~\ref{fig:usdata}d-1). And he modified the other exemplar into an orthogonal layout (Figure~\ref{fig:usdata}e-1).}

    

    

\added[id=pan]{We ensured that within the same dataset, the Weisfeiler-Lehman similarities between three target substructures and the exemplar are greater than 0.7. We recorded all modifications made by the expert along with a list of instructions (see Suppl. Material).
}

\deleted[id=pan]{We extracted 5 exemplars from these datasets (one from the Mouse-Brain dataset, one from the Email-Eu-core dataset, two from the High-School-contact dataset, and one from the Euroroad dataset, respectively). For each dataset, we extracted three target substructures whose Weisfeiler-Lehman similarities with the exemplar are larger than $0.7$. One exemplar and three targets in one dataset formed a test case, yielding 5 cases: $D_E$ was from the Email-Eu-core dataset (Figure~\ref{fig:usdata}a), $D_M$ was from the Mouse-Brain dataset (Figure~\ref{fig:usdata}b), $D_R$ was from the Euroroad dataset (Figure~\ref{fig:usdata}c), and $D_{HS1}$ and $D_{HS2}$ were from the High-School-contact dataset (Figure~\ref{fig:usdata}(d, e)). $D_E$ was used in the tutorial and the others were used in the formal study.}

\textbf{Participants and apparatus.}
Twelve volunteers were recruited to participate in the study (5 males, 7 females; aging from 23 to 27). All participants were students or researchers concentrating in computer science. They are familiar with visualization and four of them major in graph visualization. The study was conducted \replaced[id=pan]{on a PC provided by us}{on a personal computer} equipped with a mouse, keyboard, and 24-inch display. The interface was displayed within a window size of 1920 $\times$ 1080 resolution. \added[id=pan]{Parameters of the modification transfer are fixed to $\alpha=1, \beta=5, \gamma=1000, \text{ and } w=1$.}

\textbf{Study Conditions.}
We tested the performance of different fine-tuning techniques (\textbf{baseline manual}, \textbf{semi-automatic}, and \textbf{fully automatic}) on a small graph layout. Each participant was asked to \replaced[id=pan]{process three target structures in all four cases (one from the Mouse-Brain dataset, one from the Euroroad dataset, and two from the High-School-contact dataset) with three techniques}{process 4 cases (1 exemplar and 3 target structures in each case) with three techniques}, yielding \deleted[id=pan]{to} 432 ($12 \text{ participants} \times 4 \text{ cases} \times 3 \text{ targets} \times 3 \text{ techniques}$) trials.

\textbf{Procedure.} 
\replaced{The study has two stages. We first trained participants on the three manipulation modes (\textbf{baseline manual}, \textbf{semi-automatic}, and \textbf{fully automatic}). They viewed a demo video of an expert's operations using data samples extracted from the Email-Eu-core dataset (Figure~\ref{fig:usdata}a), and then practiced till they felt comfortable with the tasks. In the formal study, they were then asked to manipulate three targets' shapes to simulate the exemplar for each case using all three techniques ($4 \text{ cases } \times 3 \text{ targets } \times 3 \text{ techniques }$ in total for each participant).
For each trial, an exemplar, a modified exemplar, and a target substructure were displayed on the interface (see Suppl. Material). Participants were asked to manipulate the target substructure to simulate modifications made on the exemplar by comparing the exemplar and the modified exemplar. They could also follow printed instructions (see Suppl. Material).
With our semi-automatic method, participants were asked to specify markers first. 
One pair of markers was constructed by clicking on two nodes, one from the exemplar and one from the target substructure.
With our fully automatic method, markers were constructed automatically. Our two methods produced initial layouts that simulate the expert's modifications and participants were asked to perform the task based on initial layouts.
Parameters and initial layouts were the same for all participants.
The order of four cases, three techniques, and three targets was randomly assigned to each participant to counterbalance learning effects. After the study, participants were interviewed to give some suggestions on our approach.}
{Before the study, an expert majored in graph visualization was invited to modify the exemplar's shape according to his domain knowledge. We recorded his instructions and modifications:}

\deleted{The exemplar taken from $D_E$ was modified into a fan-like shape (Figure{~\ref{fig:usdata}} (f)).}

\deleted{The exemplar taken from $D_M$ is a star-like structure. The interior node should stay in the center of the exemplar and leaves should be evenly placed around the exemplar (Figure{~\ref{fig:usdata}} (g)).}

\deleted{The exemplar taken from $D_R$ is a circle-like structure. Its shape should be as circular as possible. (Figure{~\ref{fig:usdata}} (h)).}

\deleted{The exemplar taken from $D_{HS1}$ was modified into an orthogonal layout. The nodes should be placed in a grid-like shape. Angles among edges should be as close to 45$^{\circ}$, 90$^{\circ}$, or 180$^{\circ}$ as possible (Figure{~\ref{fig:usdata}} (i)).}

\deleted{The exemplar taken from $D_{HS2}$ contains a circle and some surrounding\deleted[id=pan]{s} nodes. The circle was modified into a regular polygon. And the edges that form the circle and surroundings nodes should be placed horizontally or vertically (Figure~\ref{fig:usdata}j).
}

\deleted{At the beginning of the study, we introduce our approach to participants and show them a tutorial with $D_E$. Then, for each case, three techniques are employed randomly. Participants need to manipulate three targets' shapes to simulate the exemplar according to the instructions given by the expert. 
We counterbalance the techniques in different participants and datasets.}


\textbf{Hypotheses.}
We measure performance by participants' completion time and number of interactions. We anticipate that the quality of the modified exemplar and the targets' layouts makes little difference because participants were asked to fine-tune the target layouts until they were satisfied.
We formulated three hypotheses:
\begin{compactenum}[\bfseries H1]
\item Our fully automatic method is more efficient than the baseline manual method.

\item Our semi-automatic method is more efficient than the baseline manual method.

\item There is no difference in performance between our semi-automatic method and our fully automatic method.
\end{compactenum}

\begin{figure}[t!p]
    \centering
    \setlength{\belowcaptionskip}{-5pt}
    \includegraphics[width=1\columnwidth]{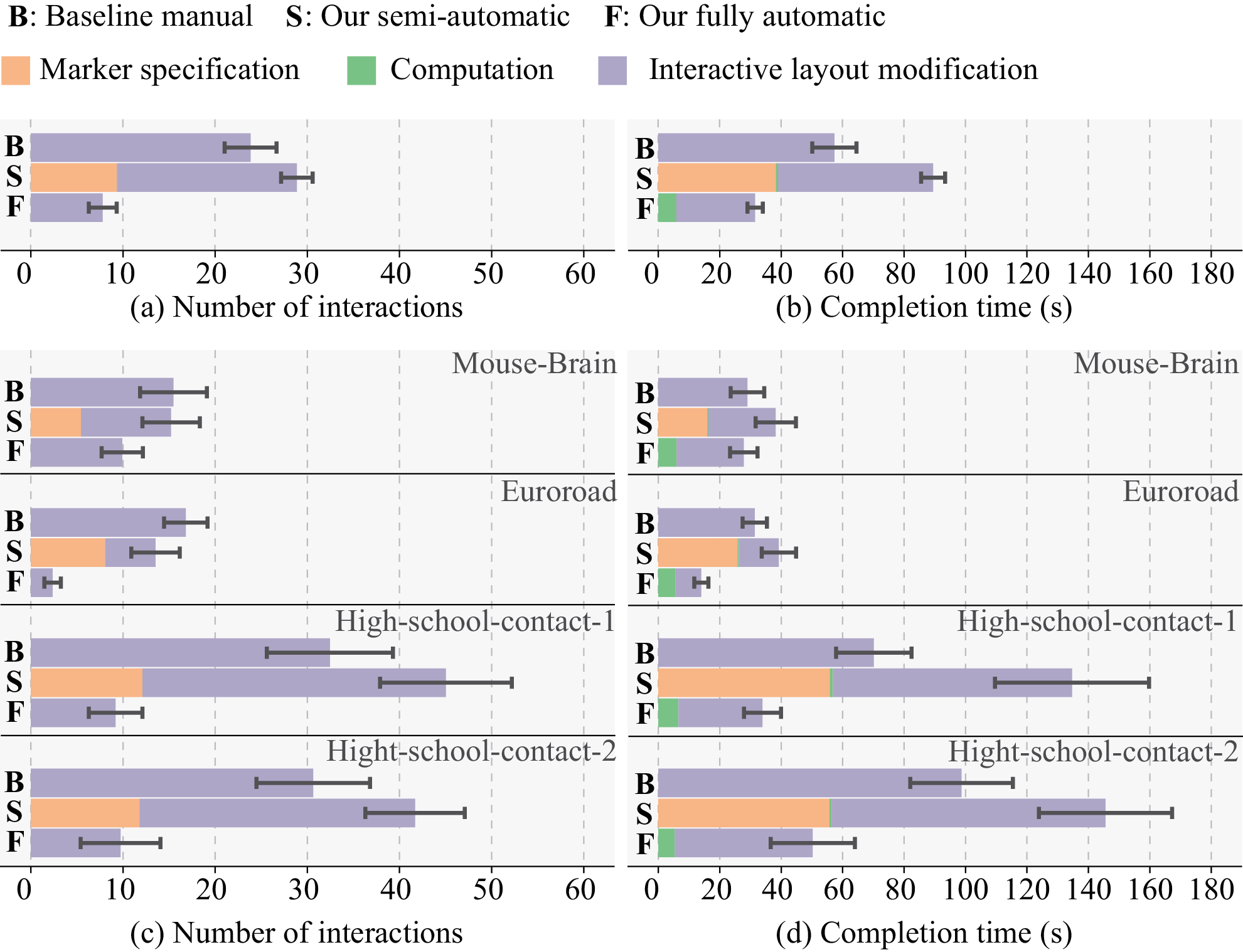}
    \caption{User study results. \added[id=pan]{Measurement components are represented as stacked bars.} (a) The distribution of the number of interactions; (b) the distribution of completion time; (c) the distribution of number of interactions on different cases; (d) the distribution of completion time on different cases. Error bars depict 95\% CIs.}
    \label{fig:usresultnew}
\end{figure}

\textbf{Results.}
\added[id=pan]{Participants spent about 45 minutes on average on the user study and got a reward of around \$5 on completion.}
We recorded the number of interactions (\added[id=pan]{mouse clicking and dragging}) that participants performed and completion times to reach a satisfying layout. The completion time includes marker specification, algorithm computation, and layout modification; and the number of interactions includes marker specification and layout modification. Figures~\ref{fig:usresultnew}(a, b) summarizes the results. 
We analyzed our results using significance tests with significance levels set to $.05$.

\added[id=pan]{The Shapiro-Wilk test, used to test the normality, suggested that both the number of interactions and the completion time did not follow normal distributions. Thus we used the Friedman test and pairwise Wilcoxon test. The Friedman test detected significant differences in both the number of interactions ($\chi^2(2)=154.96, p<0.05$) and the completion time ($\chi^2(2)=154.625, p<0.05$). Paired Wilcoxon tests were performed on all cases to compare the efficiency among three techniques. There were significant differences among all combinations of three techniques (\textbf{baseline manual}, \textbf{semi-automatic}, and \textbf{fully automatic}) on two measurements (the number of interactions and the completion time). The post-hoc analysis (Figures~\ref{fig:usresultnew}(a, b)) showed that our semi-automatic method performed most efficiently in both two measurements, followed by the baseline method and last our semi-automatic method. Thus \textbf{H1} held while \textbf{H2} and \textbf{H3} were rejected.}

\deleted{
\textbf{The number of interactions.} The Shapiro-Wilk test is employed to test the normality of the numbers of interactions. The results suggest that the numbers of interactions do not follow a normal distribution. Thus we use the Friedman test and pairwise Wilcoxon test. The Friedman test detects significant differences ($\chi^2(2)=103.10, p<0.05$). Paired Wilcoxon tests are performed on all cases to compare the efficiency among three techniques. There are significant differences among all different combinations of three techniques ($T_f$ (fully-automatic), $T_s$ (semi-automatic) and $T_m$ (manual)). The post-hoc analysis shows that $T_f$ performs most efficiently, followed by $T_s$ and $T_m$. Thus, \textbf{H1} and \textbf{H2} hold while \textbf{H3} is rejected.}

\deleted{\textbf{Completion time.} The Shapiro-Wilk test suggests that the completion time does not follow a normal distribution. Thus we use the Friedman test to test differences among three techniques. It detects significant differences among three techniques. Significant differences are detected by paired Wilcoxon tests among all different combinations of three techniques. Figure~\ref{fig:usresult} (b) shows that $T_f$ is the most efficient technique. And $T_s$ performs slightly better than $T_m$. As such, \textbf{H3} is rejected while \textbf{H1} and \textbf{H2} hold.}

\deleted{Overall, the results indicate our approaches $T_f$ (fully-automatic) and $T_s$ (semi-automatic) perform better than manually reconfiguration, which fully supports \textbf{H1} and \textbf{H2}, but mostly contradicts \textbf{H3}.}

\textbf{\added[id=pan]{Feedback.}}
\added[id=pan]{We collected some representative participant feedback. Most of them made comments along the lines of, ``\textit{In fully automatic mode, most results are pretty close to exemplar's results. I have to make little effort to modify them, especially in complex cases. But I still have to verify whether there is room for improvement}".
Many of them mentioned that they were encouraged to attempt higher quality by the high-quality result generated by the fully automatic method. Some of them mentioned that ``\textit{It is boring to wait for the fully automatic method to calculate the result}". Another complaint about our methods is that markers are hard to determine. Most participants had little experience in graph visualization. Interestingly, several participants mentioned that ``\textit{The user study is like a game, fine-tuning layouts makes me feel relaxed because I generate nice-looking results}". One of them suggested expanding our user study into an online system to collect more user data.
}

\textbf{Discussion.}
\replaced[id=pan]{
We split the number of interactions and the completion time to look for deeper insights. The completion time consists of three parts: marker specification, algorithm computation, and interactive layout modification (Figure~\ref{fig:usresultnew}d). The computation time occupies a small fraction (in green) in both the semi-automatic and fully automatic methods. The marker specification (in orange) contributes a lot to the completion time of our semi-automatic method. In most cases, participants spent most time on interactively modifying layouts.
We also calculated average completion time per interaction for the three methods; participants spent an average of 2.4 second, 2.6 second, and 3.3 second on each layout modifying interaction using the baseline method, our semi-automatic method, and our fully automatic method, respectively.
Participants spent more time thinking about and verifying results generated by our fully automatic method.
Each interaction for marker specification takes an average of 4.1 seconds. 
We observe that almost all participants tended to choose internal nodes in the star-like structures (Figure~\ref{fig:usdata}b) as markers. However, for structures extracted from the High-School-contact dataset, markers were diverse among participants.
We report results of specified markers by an expert on graph analysis in the Suppl. Material.
A good pair of markers should be able to assume the same role or status in the source and the target (e.g., cut nodes).
This indicates that experience in and knowledge of graph analysis are necessary for marker specifications.
}
{The user study verifies the efficiency of our approach. Details of results of our user study are shown in Figure~\ref{fig:usresult} (c) and (d). Our approach supports specifying markers according to their domain knowledge. The user study suggests that participants perform diversely when markers are specified by themselves (semi-automatic, $T_s$). Error bars show that $T_s$ performs more efficiently than $T_m$ for $D_M$ and $D_R$ while $T_m$ is slightly better for $D_{HS1}$ and $D_{HS2}$. The reason may be that topology structures of structures in $D_{HS1}$ and $D_{HS2}$ are more complex. Participants with little experience in graph visualization can not easily select markers.
While for $D_M$ and $D_R$ whose topological structures are more simple, $T_s$ performs better than $T_m$. Markers are more easily to be determined. Another finding is that the average completion time per interaction differs in different techniques. The average time of $T_f$ is much larger than those of others.
The reason may be that completion time of $T_f$ is mainly spent on computing correspondences. Note that, our approach employs factorized graph matching (FGM) is employed in our approach, whose time complexity for matching $S=(V^s, E^s)$ and $T=(V^t, E^t)$ is $O(k((|V^s| + |E^s|)(|V^t| + |E^t|)) + max(|V^s|^3 + |E^s|^3))$, where $k$ is the number of iterations in FGM. 
}

%% file: chapter/6.Discussion.tex
\section{Discussion \added[id=pan]{and Limitations}} \label{sec:discussion}
\replaced{In terms of the performance of modification transfer, our algorithm outperforms the baseline method (manual node dragging), as demonstrated in Section~\ref{sec:userstudy}. It reduces or eliminates the laborious interactions. And in terms of layout editing, our modification transfer algorithm may be more flexible than rule-based layout approaches~\cite{DBLP:journals/cgf/HoffswellBH18, DBLP:journals/tvcg/KiefferDMW16, DBLP:journals/tvcg/WangWSZLFSDC18}. Rather than pre-defining a set of rules or metrics, our algorithm supports arbitrary modifications on the exemplar.
}{
Our approach supports tuning multiple substructures by following a single exemplar.
It offers several advantages over existing methods. First, our approach has a broader scope than rule-based layout approaches. Rather than pre-defining a set of rules or metrics, our approach supports arbitrary modifications on the exemplar.
Second, our approach reduces or eliminates the programming workload or laborious interactions by transferring modifications automatically. Modifications made on the exemplar can be regarded as user preferences. Similar substructures can be modified by transferring the exemplar's modifications. 
}

\textbf{Usability.} 
\replaced[id=pan]{Our visualization interface is implemented with a set of fundamental interactions, such as lasso, drag, pan, and zoom. The user can easily explore the entire graph and specify substructures. Compared to box selection, lasso interaction enables the user to more freely specify a substructure with a closed path. However, for complex graphs, layout algorithms can lead to visual clutter. It is hard for the user to specify structures in a virtual plane, so that selection interactions such as filter and query will be suitable for complex cases.}{We believe our approach can reach high usability because it is implemented with a set of fundamental interactions such as Lasso, drag, pan, and zoom. However, there are some limitations. One limitation is that our approach requires the exemplar and the target to be adequately similar, which is not easy to be found. It decreases the usability. Although our approach enables substructure retrieval, its performance depends on the node embedding technique. Several embedding techniques are supported in our approach including: REGAL~\cite{DBLP:conf/cikm/HeimannSSK18}, Feature-based kernel~\cite{DBLP:journals/tvcg/ChenGHPNXZ19}, Graphlet kernel~\cite{DBLP:journals/cn/MarcusS12}, Node2vec~\cite{DBLP:conf/kdd/GroverL16}, Struct2vec~\cite{10.1145/3097983.3098061}, GraphWave~\cite{DBLP:conf/kdd/DonnatZHL18}. GraphWave has proven to be an appropriate technique in substructures retrieving~\cite{DBLP:journals/tvcg/ChenGHPNXZ19}, but it does not work well all the time because it only suggests fuzzy results. In the future, we will improve the retrieval quality to improve usability.
The user study suggests that manually manipulate a layout by dragging nodes can be laborious. Although our approach reduces the user's repetitive fine-tuning interactions, the user still has to modify an exemplar for demonstration. In the future, we plan to improve the user's performance on fine-tuning exemplars.
}

\textbf{Scalability.} Our cases show that our approach can handle fine-tuning on large-scale networks. Our interface with a WebGL rendering engine supports visualizing large-scale graphs with rich user interactions.
\added[id=pan]{Three aspects influence the scalability:}
\begin{compactenum}[\bfseries 1)]
\item \added[id=pan]{\textbf{The substructure retrieval algorithm} has a computational complexity of $O(|V^s| \times N)$, where $N$ denotes the node number of the underlying graph~\cite{DBLP:journals/tvcg/ChenGHPNXZ19}.
However, heuristic user-adjustments of the parameter $k$ (see Section~\ref{sec:retrieval}) may reduce scalability.}

\item \added[id=pan]{\textbf{Modification transfer} consists of three parts: graph matching, correspondence filtering, and two rounds of layout simulation. The time complexity of FGMU~\cite{zhou2012factorized} for matching $S=(V^s, E^s)$ and $T=(V^t, E^t)$ is $O(k \times \max(|V^t|^3, |V^s|^3) + |E^t| |E^s|^2))$, where $k$ is the number of iteration 
for FGMU. The average time complexity of correspondence filtering is $O(\min(|V^t|, |V^s|) \times |E^t| |E^s| / (|V^t| |V^s|) )$. The first round of layout simulation involves several iterations. The number of iterations depends on the number of markers. More markers can lead to less iterations. For each iteration, the deforming step employs a procedure similar to the stress-majorization layout~\cite{DBLP:conf/gd/GansnerKN04}, whose time complexity is the same as the stress majorization. The time complexity of the matching step is dominated by the Hungarian algorithm, whose complexity is $O(m^3)$, where $m$ is the number of nodes selected for matching. The second round of layout simulation runs one time because no more correspondences are built.}

\item \added[id=pan]{\textbf{The global optimization} runs as fast as the stress-majorization layout, which is sensitive to the number of nodes in the surroundings to be optimized.}
\end{compactenum}


\deleted{
Our cases show that our approach can handle fine-tuning on large scale networks. Our interface with a WebGL rendering engine is amenable for visualizing large-scaled graphs with rich user interactions. As only the surroundings of modified substructures are deformed, the computational cost of the global layout optimization is reduced. However, building correspondences can be computationally expensive. In the future, we will eliminate our approach's dependency on markers. As such, the computational cost can be decreased. Besides, we plan to accelerate our approach by implementing a GPU-based version.
}

\textbf{Robustness.} Case studies and user study indicate that our approach can handle different kinds of datasets and layouts. Our approach is not sensitive to the original layout, because we layout the exemplar and targets with the same force-directed algorithm before building correspondences.
Although the user study suggests that our \replaced[id=pan]{fully automatic method}{approach} works efficiently, we found that participants still performed a few interactions based on results generated by our approach. The reason may be that our approach generates similar layouts as the exemplar, not the same layouts; participants must check whether generated results can be improved.

\textbf{Limitations and future work.}
\added[id=pan]{This work has several limitations. 
First, the usability of the marker specification 
can be improved.
We plan to allow the user to interactively select markers from correspondences built by graph-matching algorithms. 
An algorithm that can rate the correctness of correspondences can improve its usability.
Second, we could also conduct a thorough user evaluation of readability.
We designed our method to transfer modifications among structures, and thus the readability of substructure layouts generated by our approach depends largely on the exemplar's modifications. 
Third, the substructure retrieval algorithm detects potentially similar structures using node embeddings. Its accuracy depends on the embedding technique.
}

\added[id=pan]{
In the future, we plan
to perform both lab-based control studies as well as insight-based studies in real-world settings on our prototype system to measure readability~\cite{DBLP:journals/tvcg/MarriottPWG12, DBLP:conf/apvis/NguyenHE17, DBLP:journals/tvcg/WuCASQC17},
to characterise the goals and effects, 
user perception, and insights.
}

%% file: chapter/7.Conclusion.tex
\section{Conclusion}
\replaced{We designed and evaluated}{In this work, we present} an exemplar-based \added{graph} layout fine-tuning approach that reduces human labor by transferring modifications made on an exemplar to other substructures. 
A user interface is developed to enable fine-tuning of graph layouts.
A quantitative comparison of two datasets with ground truth indicates that our approach can reach more accurate correspondences. Three case studies show that our approach works well on different datasets and layouts. A user study shows that our approach significantly reduces or even eliminates laborious interactions.